\documentclass[prd,twocolumn,nofootinbib]{revtex4-1}

\usepackage{graphicx,hyperref,verbatim,layout,textcomp}
\usepackage{subfig,mathrsfs}
\usepackage[fleqn]{amsmath}

\usepackage{placeins}
\usepackage{afterpage}
\usepackage{morefloats}
\usepackage{soul}
\usepackage{color}
\usepackage{booktabs,caption,fixltx2e}
\usepackage[flushleft]{threeparttable}
\usepackage[normalem]{ulem}

\newcommand{\twoF}{{2\cal{F}}}
\newcommand{\F}{{\cal{F}}}
\newcommand{\avgTwoF}{{\widehat{\twoF}}}

\newcommand{\apjl}{The Astophysical Journal Letters}   
\newcommand{\aap}{Astronomy and Astrophysics}   
           
\begin{document}

\title{Optimally setting up directed searches for continuous gravitational waves in Advanced LIGO O1 data}

\author{Jing Ming$^\mathrm{1,2,3}$, Maria Alessandra Papa$^\mathrm{1,2,4}$, Badri Krishnan$^\mathrm{2,3}$,  Reinhard Prix$^\mathrm{2,3}$, Christian Beer$^\mathrm{2,3}$, \\Sylvia J. Zhu$^\mathrm{1,2}$, Heinz-Bernd Eggenstein$^\mathrm{1,2}$, Oliver Bock$^\mathrm{2,3}$,  Bernd Machenschalk$^\mathrm{2,3}$}

\affiliation{$^1$Max-Planck-Institut f{\"u}r Gravitationsphysik, Albert Einstein Institute, am M{\"u}hlenberg 1, 14476, Potsdam-Golm, Germany\\
$^2$Max-Planck-Institut f{\"u}r Gravitationsphysik, Albert Einstein Institute, Callinstra{$\beta$}e 38, 30167, Hannover, Germany\\
$^3$Leibniz Universit{\"a}t Hannover, Welfengarten 1, 30167, Hannover, Germany\\
$^4$University of Wisconsin-Milwaukee, Milwaukee, Wisconsin 53201, USA
%$^5$Max-Planck-Institute for Gravitational Physics, Albert  Einstein Institute, Callinstrasse 38, 30167 Hannover
 }

\begin{abstract}

  In this paper we design a search for continuous gravitational waves
  from three supernova remnants: Vela Jr., Cassiopeia A (Cas A) and
  G347.3.  These systems might harbor rapidly rotating neutron stars
  emitting quasi-periodic gravitational radiation detectable by the
  advanced LIGO detectors.  Our search is designed to use the
  volunteer computing project Einstein@Home for a few months and
  assumes the sensitivity and duty cycles of the advanced LIGO
  detectors during their first science run.  For all three supernova
  remnants, the sky-positions of their central compact objects are
  well known but the frequency and spin-down rates of the neutron
  stars are unknown which makes the searches computationally limited.
  In a previous paper we have proposed a general framework for
  deciding on what target we should spend computational resources and
  in what proportion, what frequency and spin-down ranges we should 
  search for every target, and with what search set-up.  Here we
  further expand this framework and apply it to design a search
  directed at detecting continuous gravitational wave signals from the
  most promising three supernova remnants identified as such in the previous work.  Our optimization procedure yields broad
  frequency and spin-down searches for all three objects, at an
  unprecedented level of sensitivity: The smallest detectable
  gravitational wave strain $h_0$ for Cas A is expected to be 2 times smaller
  than the most sensitive upper-limits published to date, and our proposed 
  search, which was set-up and ran on the volunteer computing project Einstein@Home, covers a much larger frequency range. 
   
\end{abstract}

\maketitle

\section{Introduction}
\label{sec:intro}

Continuous gravitational wave (CW) signals at frequencies between
$\approx 10-1000$ Hz are expected to be emitted by rapidly rotating
compact objects with shapes that are not perfectly axially symmetric.
Because these signals are extremely weak \cite{S6BucketFUs, S6allSky2,
  knownPulsarsO1}, one needs to combine the data over very long
periods of time (months) in order to raise the signal significantly
above the average noise level. On the other hand the ability to
resolve different waveforms increases very quickly with the duration
of the observation time, and hence CW searches over broad ranges of
different waveforms are computationally very expensive.  We can
broadly characterize CW searches as targeted, all-sky and directed
according to the computational cost requirements.  

\emph{Targeted} searches for CW signals from objects like the Crab and
Vela pulsars are very inexpensive. The reason is that since their sky
position and frequency evolution are known from electromagnetic 
observations, one only needs to search for a single waveform or a
small number of waveforms around it. About 200 known pulsars rotate at
frequencies such that the expected gravitational wave (GW) signal
falls into the high sensitivity band of LIGO and Virgo.  Searches
for GW signals from these objects have been systematically carried out
throughout all observing runs
\cite{Abbott:2008fx,Abadie:2011md,Aasi:2014jln,knownPulsarsO1}.

At the other end of the spectrum of possible searches are the
so-called all-sky searches where one has no prior information on any
specific source. Typically these searches span broad ranges of signal
frequencies, tens or hundreds of Hz, source positions anywhere in the
sky and spindown parameter ranges varying by up to two orders of
magnitude \cite{VSR4LowFreq,S6allSky2,S6BucketFUs,O1AS20-100}.
%The parameter space we need to search would be very large, so in this case, the comp}uting resources are mostly limited. Some blind searches have also been carried out in\cite{S2ScoX1,S2Hough,S4PSH,S5Powerflux2009,S5Powerflux2011, EatHS4R2,EatHS5R1,Abbott:2008uq,s6allsky, S6allSky2} and some of them \cite{EatHS4R2,EatHS5R1,Abbott:2008uq, s6allsky} have been performed on the public distributed c\cite{OptimalMethod}uting project Einstein@Home \cite{Einstweb}.  
  
Somewhere in between these two extremes lie the \emph{directed}
searches in which we look for signals from an interesting sky region
or sky point, e.g. Cas A \cite{S6CasA} and generally young supernova
remnants \cite{S6NineYoungSNRs}, the Orion Spur \cite{S6Orion},
Globular cluster NGC 6544 \cite{S6NGC6544}, the low mass x-ray
binaries Scorpius X-1 and XTE J1751-305
\cite{MeadorsS6TargetedBinary}, and the Galactic center
\cite{GalacticCenter2013}. In these searches the computational cost
can still be very high because many different waveforms may have to be
searched for, corresponding to different signal phase evolutions.

Even for supernova remnants which are very well localized (i.e. for
which the localization uncertainty is smaller than the sky resolution
of the GW search), the breadth of different waveforms that
we would need to search over, corresponding to
different frequency and spindown parameters values, is large enough
that a single coherent search is computationally unfeasible.  We
resort then to semi-coherent search schemes
\cite{HierarchP1,HierarchP2,Papa:2000wg,HoughP2,
  HierarchP3,Pletsch:2009uu,GlobCorr}. In these, the data set is split
into $N$ shorter segments of duration $T_{\mathrm{coh}}$.  Each of
these segments is matched with signal templates coherently and in the
end the results from these segments are combined incoherently. By
reducing the coherent duration $T_\mathrm{coh}$, at fixed mismatch,
the resolution of the waveform parameters becomes coarser and the
computing cost is reduced. But this comes at the cost of a reduced
sensitivity. In realistic scenarios, we do not have unlimited
computational power available to us.  At fixed computing power and for
a given data set, there are trade-offs to be made between the length
of the coherent segment duration, the resolution in the waveform
template bank (the mismatch) and the breadth of the parameter space
that we search.  Furthermore, these trade-offs should also fold in
what we know about the targeted objects, how far they are and how old
they are, and the priors that we hold on their deformation and
rotation frequency. In computationally limited problems like the ones
we are dealing with here, these trade-offs are critical; making the
wrong choices could lead us to miss detectable signals and 
waste significant computational resources.

In \cite{OptimalMethod} we presented an optimization method which, for
a given computational budget, a choice of astrophysical priors, and
the measured computational efficiency of the search software, tells us
how to distribute computing resources for directed CW searches by
maximizing the detection probability.  The search parameters
(including, but not limited to, $T_{\mathrm{coh}}$ and $N$) which will be
called \emph{search set-ups}, determine both the computational cost
and sensitivity of the search.  The optimization method determines the
optimal search set-ups for the different targets at different waveform
frequencies. The basic idea is to break the large parameter space of
the allowed gravitational signals into many small \emph{cells}, and to
estimate the computing cost and the detection probability for each
cell.  Note that the computing cost for each cell depends on
parameters related to the search pipeline and software.  In particular
it depends on the aforementioned coherent segment duration
$T_{\mathrm{coh}}$, the spacing of the template bank (these are part
of the search ``set-up''), and also the computational efficiency of
the software.
%Constrained by the total amount of available computing resources the whole parameter space can't be searched, so 
We use the detection probability and computational cost to rank what
cells and with what search set-up should be searched so that the
overall detection probability is maximized. This is done with a linear
programming technique. The interplay between the different factors is
not trivial and the results are not easy to guess without using the
optimization method.

%% \jingcomment{A  previous work \cite{Prix:2012yu} used to optimize the  set-up of the S6 Cas A search \cite{S6CasA} is just fixed the parameter space.}
A semi-analytic optimization scheme \cite{Prix:2012yu} was previously used to design a directed search for continuous signals from Cas A \cite{S6CasA}. Such a scheme minimizes the smallest detectable signal at fixed parameter space and fixed computing cost. In comparison to that scheme, the advantage of the optimization approach \cite{OptimalMethod} that we use here, is that not only does it provide the optimal search set-up, but it also tells us where in parameter space we should spend the computing  budget and on what targets. Furthermore the astrophysical priors are explicit, forcing us to spell out what they are. In previous schemes, including \cite{Prix:2012yu}, the interplay between different priors is often folded-in in a priori choices that are not transparent. %volumesallows in addition for optimally selecting the
%best targets and most promising parts of parameter space to search,
%while the method in \cite{Prix:2012yu} is restricted to a fixed given
%parameter space.

In this paper we present the first application of \cite{OptimalMethod} to set
up a directed search by using the first observational run (O1) data of the
Advanced LIGO detectors \cite{2016ligo}.  Following
\cite{OptimalMethod}, we investigate different astrophysical targets
and priors and consider a range of search set-ups for different
coherent segment durations, and optimize over all these. We expand
with respect to \cite{OptimalMethod} because i) whereas in
\cite{OptimalMethod} for each coherent segment duration we considered
a single grid spacing combination, here we consider for each coherent
segment duration many different combinations of grid spacings and
optimize over these, ii) for each grid spacing combination, we fold in
the {\it {measured}} mismatch distributions obtained from the existing
search codes for all the grid spacings and set-ups considered, rather
than the analytical estimate which can have very large errors, iii) we
revise the computational cost model for our search software to account
for recent enhancements in computational efficiency in the search software, iv) we evaluate
the loss in detection efficiency incurred if our estimate for the age
of a target object is wrong, and finally v) we introduce
simplifications with respect to the strictly optimal solution based on
practical considerations and estimate the impact on the detection
probability.

The plan of this paper is as
follows: %We start with the scientific motivation of the search in Sec.~\ref{sec:mt}.
in Section \ref{sec:signal} we recall the basics of the signal that we
want to detect, and introduce quantities that will be referred to in
the rest of the paper. The search that we want to perform is
introduced in Section \ref{sec:TheSearch}, followed by Section \ref{subsec:astroprior} where we discuss the astrophysical priors on the signal population. In Section 
\ref{sec:OpSetup} we carry out the optimization for the three
supernova remnants and finally, in Section \ref{sec:conclusions}, we summarize and discuss 
the main results.

\section{The expected gravitational wave signal}
\label{sec:signal}

For any plane gravitational wave in standard general relativity, we
can choose a wave frame transverse to the direction of propagation
such that the two polarizations have the form
\begin{eqnarray}
h_+ (t)  =  A_+ \cos \Phi(t) \nonumber \\
h_\times (t)  =  A_\times \sin \Phi(t).
\label{eq:monochromatic}
\end{eqnarray}
The phase $\Phi(t)$ of the waves that we target with LIGO is a rapidly varying function of time while the amplitudes $A_{+,\times}$ are generally slowly varying. In fact, consider 
a rapidly rotating neutron star and let $\iota$ be the angle between
the total angular momentum of the star and the direction from the star
to Earth.  The amplitudes of the signals of
Eq.~(\ref{eq:monochromatic}) are constant over time:
\begin{eqnarray}
A_+  & = & {1\over 2} h_0 (1+\cos^2\iota) \nonumber \\
A_\times & = &  h_0  \cos\iota. 
\label{eq:amplitudes}
\end{eqnarray}
Here $h_0$ is the intrinsic gravitational wave amplitude.  The signal
at the detector, $h(t)$, is a superposition of the two polarizations
\begin{equation}
h(t)=F_+ (\alpha,\delta,\psi ;t) h_+ (t) + F_\times (\alpha,\delta,\psi; t) h_\times(t),
\label{eq:signal}
\end{equation}
where $F_+ (\alpha,\delta,\psi;t)$ and
$F_\times(\alpha,\delta,\psi;t)$ are the detector beam pattern
functions for the two polarizations. Here $(\alpha,\delta)$ are the
right-ascension and declination for the source and $\psi$ is the
orientation of the wave-frame with respect to the detector frame. Due
to Earth's rotation around its axis and its orbit around the Sun, the
relative orientation between the detector and the source is changing
continuously which makes $F_{+,\times}$ time varying.

$\Phi(t)$ is the phase of the gravitational wave signal at time
$t$. If $\tau_{\mathrm{SSB}}$ is the arrival time of the wave with
phase $\Phi(t)$ at the solar system barycenter, then the phase has the
form
\begin{multline}
\label{eq:phiSSB}
\Phi(\tau_{\mathrm{SSB}}) = \Phi_0 + 2\pi [ f_0 (\tau_{\mathrm{SSB}}-{\tau_0}_{\mathrm{SSB}})  +
\\ {1\over 2} \dot{f}_0 (\tau_{\mathrm{SSB}}-{\tau_0}_{\mathrm{SSB}})^2 + {1\over 6} \ddot{f}_0 (\tau_{\mathrm{SSB}}-{\tau_0}_{\mathrm{SSB}})^3 + \cdots ].
\end{multline}
The transformation between detector time $\tau_{\mathrm{De}}$ and SSB time
$\tau_{\mathrm{SSB}}$ is
\begin{equation}
 \label{eq:tau(t)} \tau_\mathrm{SSB}(\tau_{\mathrm{De}}) = \tau_{\mathrm{De}} + \frac{\mathbf{r}(\tau_{\mathrm{De}}) \cdot \mathbf{n}}{c} +
\Delta_{\mathrm{E}\odot} - \Delta_{\mathrm{S}\odot}\,,
\end{equation} 
where $\mathbf{r}(\tau_{\mathrm{De}})$ is the position vector of the detector in the SSB
frame, $\mathbf{n}$ is the unit vector pointing to the source,
and $c$ is the speed of light; $\Delta_{\mathrm{E}\odot}$ and $\Delta_{\mathrm{S}\odot}$
are respectively the relativistic Einstein and Shapiro time
delays.

\section{The Search} 
\label{sec:TheSearch} 

Based on the results of \cite{OptimalMethod}, we concentrate on three
targets, Vela Jr., Cas A and G347.3, and illustrate the design of three
directed searches for continuous GW signals from these.
These three targets are supernova remnants and are believed to harbour a neutron star. The electromagnetic observations identify these as point sources. The position of Vela Jr. comes from Chandra X-ray satellite data \cite{2001ApJ...559L.131P}.
The estimates of its age and distance are uncertain \cite{owen2014}. Cassiopeia A (Cas A in short) is one of the youngest known supernova remnants. The position of Cas A comes from Chandra data \cite{1999IAUC.7246....1T}, the distance from \cite{1995ApJ...440..706R} and the age from \cite{2006ApJ...645..283F}. G347.3 is a target which is not only close (1.3 kpc \cite{2004A&A...427..199C}) but also young (1600 years old \cite{1997A&A...318L..59W}). The position of G347.3 is also from Chandra data \cite{2008A&A...484..457M}. 

We assume
using the volunteer computing project Einstein@Home \cite{Einstweb}, data with the
average noise of the Advanced LIGO detectors during their first
observational run (designated as O1), and with the duration and average duty
factor of the LIGO O1 data.

\subsection{Search method} 
\label{method1}

As mentioned earlier, we take a semi-coherent approach in which the
data is divided into shorter segments each spanning the same
observation time $T_\mathrm{coh}$. Each of these segments are analyzed
coherently and afterwards combined incoherently. The coherent analysis
in each segment uses the $\mathcal{F}$ statistic introduced in
\cite{JKSPaper,MultiIfoFstat}.  The final semi-coherent detection
statistic is the average of the $\mathcal{F}$-statistic for each
segment which shall be denoted as $\avgTwoF$ as implemented in the GCT
(Global correlation transform) method \cite{Pletsch:2009uu}, used in
the most recent Einstein@Home papers \cite{S6allSky2,S6BucketFUs,
  S6CasA}.
%hierarchical approach of GCT (Global correlation transform) method \cite{Pletsch:2009uu} 
%This method is known as semi-coherent method.  
%In a  semi-coherent search, the final result is obtained in two steps. The first is the coherent step in which each segment is searched with a matched filtering method and afterwards the statistic $\F_\ell$ for the $\ell^{th}$ segment  is obtained \cite{MultiIfoFstat}. The second step is called incoherent step in which the  $\F_\ell$ are combined in a certain way. Here we consider a stack-slide type to combined $\F_\ell$ across the $N$ segments: 
%\begin{equation}
%  \label{eq:15}
%  \avgF = \frac{1}{N}\sum_{\ell=1}^N\F_\ell\,.  
%\end{equation}
%After summed up $\F_\ell$ and then divided by number of segments, the $\avgF$ is obtained in stack-slide incoherent step. 
% In Gaussian noise,  $\avgTwoF$ follows a non-central chi-squared distribution
% with $4N$ degrees of freedom \cite{JKSPaper}.

\subsection{Template banks and mismatch distributions}
\label{sec:templateBanks}

As mentioned earlier, each waveform is defined by the parameters
($h_0,f,\dot{f}, \ddot{f},\alpha,\delta, \psi,\iota,\Phi_0$).  However
the coherent search method analytically maximizes over the parameters
($h_0,\psi,\iota,\Phi_0$) and so we only need to explicitly search
over ($f,\dot{f},\ddot{f}, \alpha,\delta$). Since we consider directed
searches the sky position is known and the different waveforms are
determined by only ($f,\dot{f},\ddot{f}$).

With the term ``template bank'' we indicate the collection of
waveforms that we explicitly search for. These are defined by values
of the waveform parameters ($f,\dot{f},\ddot{f}$). The spacings in
these parameters in principle has to be fine enough that a real
signal, with waveform parameters lying between adjacent points in the
template bank would still be detectable.  

Two grids are set-up: a coarse grid, used for the coherent searches,
and a fine grid, used for the incoherent searches. The frequency
spacing $\delta f$ is the same for both the coherent and incoherent searches
whereas the frequency derivative spacings are refined by a factor of
$\gamma^{(1)}$ and $\gamma^{(2)}$ going from the coherent to the
incoherent grids:  $\delta \dot{f}_f={\delta \dot{f}_c \over \gamma^{(1)}}$ and $\delta \ddot{f}_f={\delta \ddot{f}_c \over \gamma^{(2)}}$. The total number of waveforms that are searched is
$N_\mathrm{fine}=\gamma^{(1)}\gamma^{(2)} N_\mathrm{coh}$ with
$N_\mathrm{coh}=N_{f_c} N_{\dot{f_c}} N_{\ddot{f_c}}$, where $N_{f_c}$,
$N_{\dot{f_c}}$ and $N_{\ddot{f_c}}$ are the number of coarse grid points in $f$,
$\dot{f}$ and $\ddot{f}$ respectively. 

The grid spacings
$\delta f_c, \delta \dot{f}_c, \delta\ddot{f}_c$ are parametrized by the parameters $m_f$,
$m_{\dot{f}}$, $m_{\ddot{f}}$ as follows \cite{Pletsch:2010xb}:
\begin{eqnarray}
\label{eq:gridSpacings}
  \delta f_c&=& \frac{\sqrt{12m_f}}{\pi T_\mathrm{coh}}\,,\\ 
  \delta\dot{f}_c &=& \frac{\sqrt{180m_{\dot{f}}}}{\pi {T_\mathrm{coh}}^2}\,,\\
  \delta\ddot{f}_c &=& \frac{\sqrt{25200 m}_{\ddot{f}}}{\pi {T_\mathrm{coh}}^3}\,.
\end{eqnarray}
%In the incoherent combination, the grids used in spin-down are finer than the coarse grids in coherent stage and the frequency grid is the same for both stages. The spacings of those fine grids are given by:
%\begin{equation}
%  \delta f^{(k)}_f = \frac{\delta f^{(k)}_c}{\gamma^{(k)}}\,,\quad k = 1,2\ldots\, 
%\end{equation}
%The semi-coherent grid spacing in each dimension has a factor $\gamma^k$ finer than the coarse grid one ($f^{(1)} = \dot{f}$ and $f^{(2)} = \ddot{f}$).
% and these refinement factors depend on the number of coherent segments:
%\begin{eqnarray}
%  \gamma^{(1)} &=& \sqrt{5N_{\mathrm{seg}}^2-4}\,,\\
%  \gamma^{(2)} &=& \frac{\sqrt{35N_{\mathrm{seg}}^4-140N_{\mathrm{seg}}^2+108}}{\sqrt{3}}\,.
%\end{eqnarray}

Every template bank for a given search set-up ($T_{\mathrm{coh}}$ and $N$) is characterized by its mismatch distribution. The mismatch $\mu$ is the quantity that measures how much signal-to-noise may be lost due to the mismatch between a signal parameters and the discrete template bank. The finer the template bank is, in general the smaller the mismatch. The mismatch distribution can be measured by simulating signals (no noise) and measuring the signal-to-noise ratio  $\rho^2_{\textrm{no-mismatch}}$ associated with a search performed with a perfectly matched template (i.e. a template with parameters identical to the signal parameters) and the signal-to-noise ratio $\rho^2_{\textrm{mismatch}}$ associated with the maximum of the detection statistic from a search performed with the template grid that we want to characterize:
\begin{equation}
 \mu:={ { \rho^2_{\textrm{no-mismatch}} -  \rho^2_{\textrm{mismatch}}}\over {{\rho^2_{\textrm{no-mismatch}}}} }.
\label{eq:mismatch}
\end{equation}
The signal-to-noise ratio $\rho^2$ for a semi-coherent $\F$-statistic search with $N$ segments is connected to the expectation value of $\hat\F~$ as follows: $E[2\hat\F] = \rho^2 + E[n]$, where $E[n]$ is the expected value of noise alone.

\begin{figure}
\centering
\includegraphics[width=0.45\textwidth]{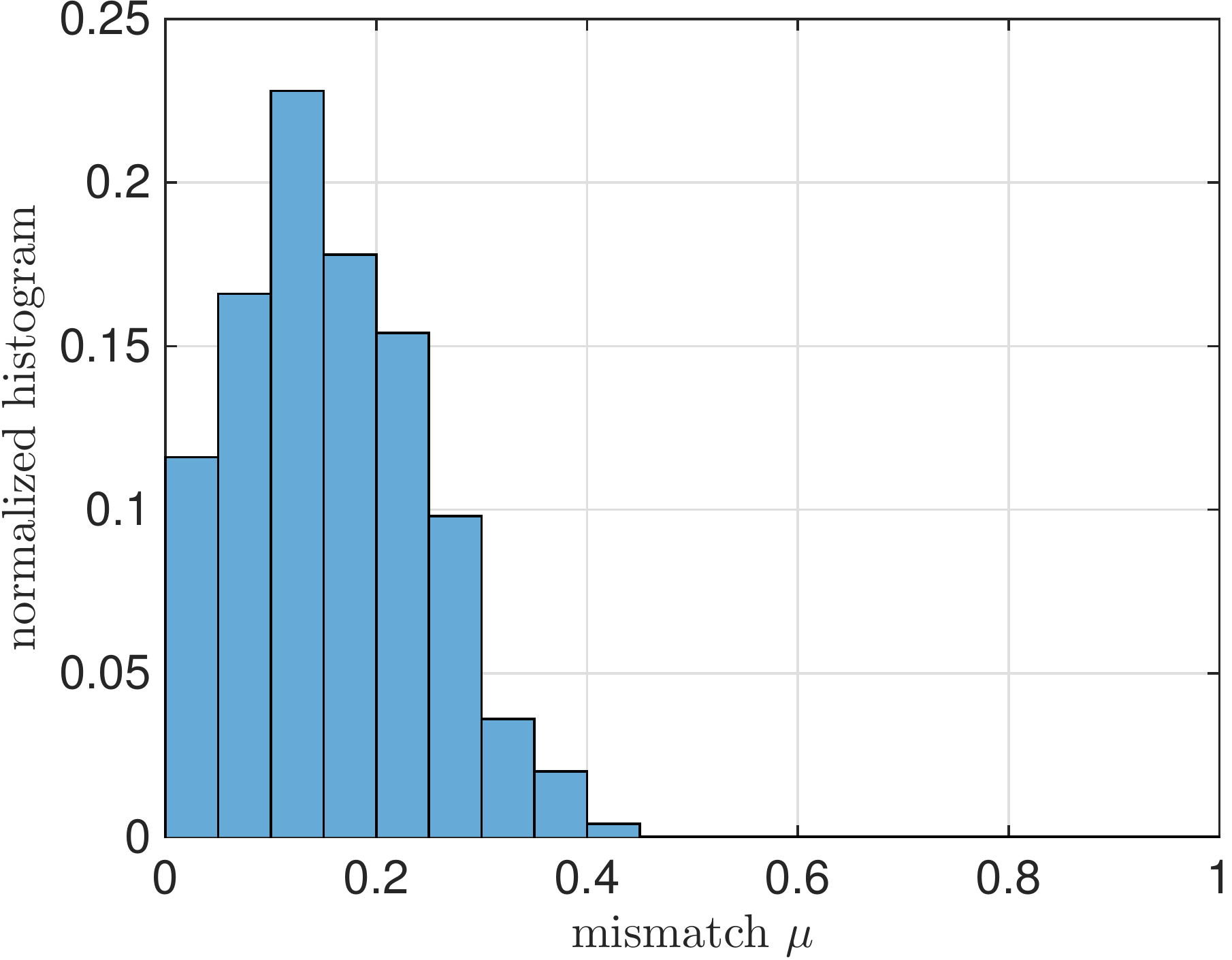}
\caption{This histogram shows the mismatch distribution for the grid spacing: $m_f=0.15$, $m_{\dot{f}}=0.3$, $m_{\ddot{f}}=0.003$, $\gamma^{(1)}=8$, $\gamma^{(2)}=20$ and $T_\mathrm{coh}$=15 days for Vela Jr and it is the set-up that was eventually chosen for the search. The average measured mismatch is $\langle \mu\rangle = 15.8\%$.}
 \label{mismatch_hist}
\end{figure}

We derive the mismatch distribution for a given search set-up $T_{\mathrm{coh}}$, $N$ and a template bank defined by $m_f, m_{\dot{f}}, m_{\ddot{f}}, \gamma^{(1)}$ and $\gamma^{(2)}$ by injection-and-recovery Monte Carlos following Eq.~(\ref{eq:mismatch}).
%In particular, 500 artificial signals are added to two Gaussian noise data streams simulating the LIGO O1 data from the two detectors. The fake noise has gaps simulating a realistic duty factor of the instruments\footnote{The fake data is created with a code \texttt{lalapps\_Makefakedata\_v4} which is a part of the LIGO Algorithm Library (LALSuite) \cite{lalSuite}.}. 
In particular, 500 artificial signals are produced with gaps simulating the realistic output from  the LIGO O1 detectors. These data streams are jointly searched with our standard semi-coherent search\footnote{The artificial data
is created with \texttt{lalapps\_Makefakedata\_v4}. The search is performed using \texttt{lalapps\_HierarchSearchGCT} (GCT). Both programs are part of the LIGO Algorithm Library (LALSuite) \cite{lalSuite}.}. Because these injection-and-recovery Monte Carlos are performed on the noise-free data streams, the computed values of the search code correspond to the expectation values of the statistic, $E[\hat{\F}]$, and therefore we obtain the mismatches directly from Eq.~(\ref{eq:mismatch}).
%The data is searched\footnote{The search is performed using \texttt{lalapps\_HierarchSearchGCT} (GCT), also a part of LALSuite.} with the given grid spacings and $T_{\mathrm{coh}}$ and $E[\hat\F_{\mathrm{meas}}]$ is obtained.  We repeat the search with a template that matches the signal parameters exactly and obtain $E[\hat\F_{\mathrm{max}}]$. Repeating this procedure for $500$ signals and following the Eq.~(\ref{eq:mismatch}), we determine the mismatch distribution. 
The range of the parameters of  the $500$ signals is given in Table~\ref{tab:missmatchparams}. We have considered over 2000 different spacings. The ranges for these spacings are listed in Table~\ref{tab:spacings}. For illustration purposes, Fig.~\ref{mismatch_hist} shows the mismatch distribution for $T_\mathrm{coh}= 15 ~{\mathrm{days}}$, $m_f=0.15$, $m_{\dot{f}}=0.3$, $m_{\ddot{f}}=0.003$, $\gamma^{(1)}=8$ and $\gamma^{(2)}=20$. The average measured mismatch is $\langle \mu\rangle = 15.8\%$ and this is a typical value for a deep broad parameter space CW survey (cfr. for instance with the average mismatch of the last stage follow-up in \cite{S6BucketFUs} which lies at $\sim$ 0.13). 

\begin{table*}
\begin{threeparttable}
\caption{
  \label{tab:missmatchparams}
 Injection parameters used in mismatch investigation}
  \begin{centering}
  \begin{tabular}{lll}
    \tableline
   
Parameter & \phantom{.} & Range
    \\
    \tableline
	%Sky position [rad] & & $\lvert\Delta\alpha\rvert + \lvert\Delta\delta\rvert \leq
%10^{-3}\,\text{rad}$ \\
	%&& with $\Delta\{\alpha,\delta\} = \{\alpha,\delta\}_\text{inj} - \{\alpha,\delta\}_\text{GC}$\\
	%Signal strength & & $h_0^\text{injected} = 4$ (signal only)\\
	Frequency [Hz] & &  $151\,\mathrm{Hz} \leq f \leq 152\,\mathrm{Hz}$\\
	First spin-down [Hz/s] & &$-10^{-7}\,\mathrm{Hz/s} \leq \dot{f} \leq -10^{-12}\,\mathrm{Hz/s} $\\
	Second spin-down [$\mathrm{Hz}/\mathrm{s}^2$] & &$ 10^{-22}\,\mathrm{Hz}/\mathrm{s}^2 \leq \ddot{f} \leq$ {$7\times10^{-17}\,\mathrm{Hz}/\mathrm{s}^2$}\\
%	Polarization angle & & $0\le \psi \le 2\pi$\\
%	Initial phase constant & & $0\le \phi_0 \le 2\pi$\\
%	Inclination angle & & $-1\le \cos\iota \le 1$\\
    \tableline
  \end{tabular}

\end{centering}
    \begin{tablenotes}
      \small
      \item\textbf{Notes:} Parameters of the fake signals used to derive the mismatch distributions. $f$ is uniformly randomly distributed; $\dot{f}$ and $\ddot{f}$ are log-uniformly randomly distributed. %Note that we use ``signal only" model in \texttt{lalapps\_Makefakedata\_v4} to avoid the influence due to noise fluctuation. Under this model, the noise strength is fixed to constant 1.
    \end{tablenotes}
\end{threeparttable}

\end{table*}

%%%%%%%%%%%%%%%table
\begin{table}
\begin{threeparttable}
\caption{
  \label{tab:spacings}
  Template bank parameters}
  \begin{centering}
  \begin{tabular}{lllll}
    \tableline
   
Parameter & \phantom{.} & Range
    \\
    \tableline

	$T_\mathrm{coh}$ [days] & & 10, 15, 20, 30, 60\\
	$m_f$& &  $ 0.1 \leq m_f \leq 1.0$\\
	$m_{\dot{f}}$& &$ 0.1 \leq m_{\dot{f}} \leq 1.0$\\
	$m_{\ddot{f}}$& &$ 0.001 \leq m_{\ddot{f}} \leq 1.5$\\
	$\gamma^{(1)}$& &$ 1 \leq \gamma^{(1)} \leq 50$\\
	$\gamma^{(2)}$& &$ 5  \leq\gamma^{(2)} \leq 100$\\

    \tableline
  \end{tabular}

\end{centering}
    \begin{tablenotes}
      \small
      \item\textbf{Notes:} The total observation duration $T_\mathrm{obs}$ is 120 days. It is divided into $N\approx T_\mathrm{obs}/T_\mathrm{coh}$ coherent segments, the ``$\approx$" due to there being gaps in the science-quality data.
    \end{tablenotes}
\end{threeparttable}

\end{table}

%%%%%%%%%figure%%%%%%%%%

\subsection{Search software timing}
\label{sec:timingModel}

In computationally limited problems, one needs a rational process for 
allocating computational resources amongst different competing
proposals. For example, should we allocate equal resources for each of
Vela Jr., Cas A and G347.3?  If not, then what is the optimal
distribution of computing power?  

An important ingredient in the optimization is an
accurate estimation of the run-time of the search pipeline for a given
search set-up over a given parameter space.  CW searches are the most
computationally expensive gravitational wave searches and a great effort has been employed to optimize them. As a result, a well developed timing model has been developed for our main  search pipeline.  Furthermore, the use of
Einstein@Home demands that we are able to predict the run-time of
the work-units assigned to each of the host machines, which adds a further
incentive to characterize the software accurately. 
 
The time $\tau_\mathrm{total}$ that it takes to
perform a search in a parameter-space volume covered by $N_\mathrm{coh}$
coarse grid templates and $N_\mathrm{fine}$ fine grid templates, using
data from $N_\mathrm{det}$ detectors, divided among $N$
segments can be written as:
\begin{equation}
\begin{aligned}
 \tau_\mathrm{total}= & NN_\mathrm{det}N_\mathrm{coh}\tau_\mathrm{RS}+NN_\mathrm{inc}\tau_{\mathrm{sum}\F}\\
&+N_\mathrm{inc}\tau_\mathrm{Bayes}+N_\mathrm{can}\tau_\mathrm{Recalc}.
\end{aligned}
\label{eq:timing}
\end{equation}
The timing coefficients $\tau_\mathrm{RS}$, $\tau_{\mathrm{sum}\F}$,
$\tau_\mathrm{Bayes}$ and $\tau_\mathrm{Recalc}$ are determined based
on computing time measurements executed with various search set-ups on
\texttt{Intel Xeon E3-1231 v3} CPUs
\footnote{http://ark.intel.com/de/products/80910/Intel-Xeon-Processor-E3-1231-v3-8M-Cache-3\_40-GH}.

The first term in Eq.~(\ref{eq:timing}) is the cost of the coherent
step.  When searching over several thousand signal frequency bins
corresponding to the same sky position and spindown values, the best
algorithmic implementation of the $\F$-statistic is obtained by
carrying out the frequency demodulation of the signal by resampling a
down-sampled time-series according to $\tau_{\mathrm{SSB}}$ and then
performing an FFT \cite{JKSPaper, 2010resampling}. Our most recent
enhancement of the search codes uses this resampling + FFT method.
$\tau_\mathrm{RS}$ is the time that it takes to calculate a value of
the coherent detection statistic, corresponding to a single template,
using data from a single detector and from a single segment. It can be
written as \cite{Reinhard_tm} :
\begin{equation}
\tau_\mathrm{RS}=\tau_\mathrm{Fbin}+{N_\mathrm{samp}^{\mathrm{FFT}}\over N_{\mathrm{Fbin}}} (\tau_\mathrm{FFT}+R \tau_\mathrm{spin}),
\label{eq:timing2}
\end{equation}
where $\tau_\mathrm{Fbin}$ is the time spent on operations on each
output frequency bin, $\tau_\mathrm{FFT}$ is the time spent on the FFT
per sample $N_\mathrm{samp}^{\mathrm{FFT}}$ of the resampled and
zero-padded time series, and $\tau_\mathrm{spin}$ is the time spent
per sample of the SSB-frame resampled time series without
zero-padding, with length $R\,N_\mathrm{samp}^{\mathrm{FFT}}$ where
$R \le 1$.  Zero-padding is used in order to obtain the desired
frequency resolution of the resulting $\F$-statistic.
The timing coefficients of Eq.~(\ref{eq:timing2}) are:  
% \repradd{[comment: on which hardware?]}
$\tau_{\mathrm{Fbin}}=6.0\times10^{-8}$ s,
$\tau_{\mathrm{FFT}}=3.3\times10^{-8}$ s,
$\tau_{\mathrm{spin}}=7.5\times10^{-8}$ s.
%However, $\tau_\mathrm{RS}$ is also set-up dependent (through $T_{\mathrm{coh}}$ and $\delta f$). Hence, giving the set-up and the above three basic timings, we can obtained  the $\tau_\mathrm{RS}$. 

The second term in Eq.~(\ref{eq:timing}) is the cost of the
  incoherent step: For every fine grid point we sum $N$
detection statistic
values. %3The c\cite{OptimalMethod}uting cost of this operation is $\tau_{\mathrm{sum}\F}$.
For $2 \leq {N}\leq 12 $ the timing coefficient
$\tau_{\mathrm{sum}\F}$ is
\begin{equation}
\tau_{\mathrm{sum}\F}= 7.28\times10^{-9} - { {3.72\times10^{-10}} {N}  }\,\,  (\mathrm{s})\,\,\,.
\label{eq:timing3}
\end{equation}
Since the efficiency of adding up detection statistic values
increases with the number of segments, there is a negative
term in Eq.~(\ref{eq:timing3}) proportional to number of segments $N$. Note that 
%\jingrep{
Eq.~(\ref{eq:timing3}) has been measured by timing the search on set-ups where $N$ varies between 2 and 12, which is the range of interest for our data set, hence, it may not hold for $N$ values outside of this range. Eq.~(\ref{eq:timing3}) has been obtained from the linear fitting of 23 timing trials with a norm of residuals $2.4\times10^{-9}$.
%}
%These timing trials are from the search on set-ups where $N$ varies between 2 and 12. Hence, relation (\ref{eq:timing3}) may not hold outside of this range.}

As done in \cite{S6allSky2,S6BucketFUs} the main detection statistic
is augmented with variants that are robust to detector artefacts,
namely the line-robust $\hat{B}_{\mathrm{{S/GL}}}$ and the
transient-robust $\hat{B}_{\mathrm{{S/GLtL}}}$, as well as a detection
statistic which is sensitive to some types of transient signals,
$\hat{B}_{\mathrm{{tS/GLtL}}}$ \cite{2014PhRvD..89f4023K,
  2015CQGra..32c5004K, Keitel2016}. The third term in
Eq.~(\ref{eq:timing}) is the time to compute these specialized
statistics {\it {given}} the single detector and multi detector coherent
detection statistic values (see Eq.~(13) of \cite{Keitel2016}). This
is the reason why $\tau_\mathrm{Bayes}$ is independent of the number
of segments. { $~\tau_\mathrm{Bayes}=4.4\times10^{-8}$ s is also obtained from the 23 timing trials}. For set-ups
with just a few segments, the cost of computing these various statistics can be larger than the cost of the incoherent step.

The last term in Eq.~(\ref{eq:timing}) is the computing cost for the
recalculation all these detection statistics at the exact fine grid
point in all the coherent segments. This is done only for the
$N_\mathrm{can}$ candidates that are in the top
list\footnote{The top list is the list of top candidates that is returned by the Einstein@Home volunteer computer to the main Einstein@Home server. Typically multiple top lists will be returned, each ranked according to a different detection statistic.}. In the last few
Einstein@Home searches the number of candidates in the top-list is
$N_\mathrm{can}=O(1000)$ and this recalculation cost is negligible
with respect to the costs of the other three terms.

\begin{figure}
\begin{centering}
\includegraphics[width=0.50\textwidth]{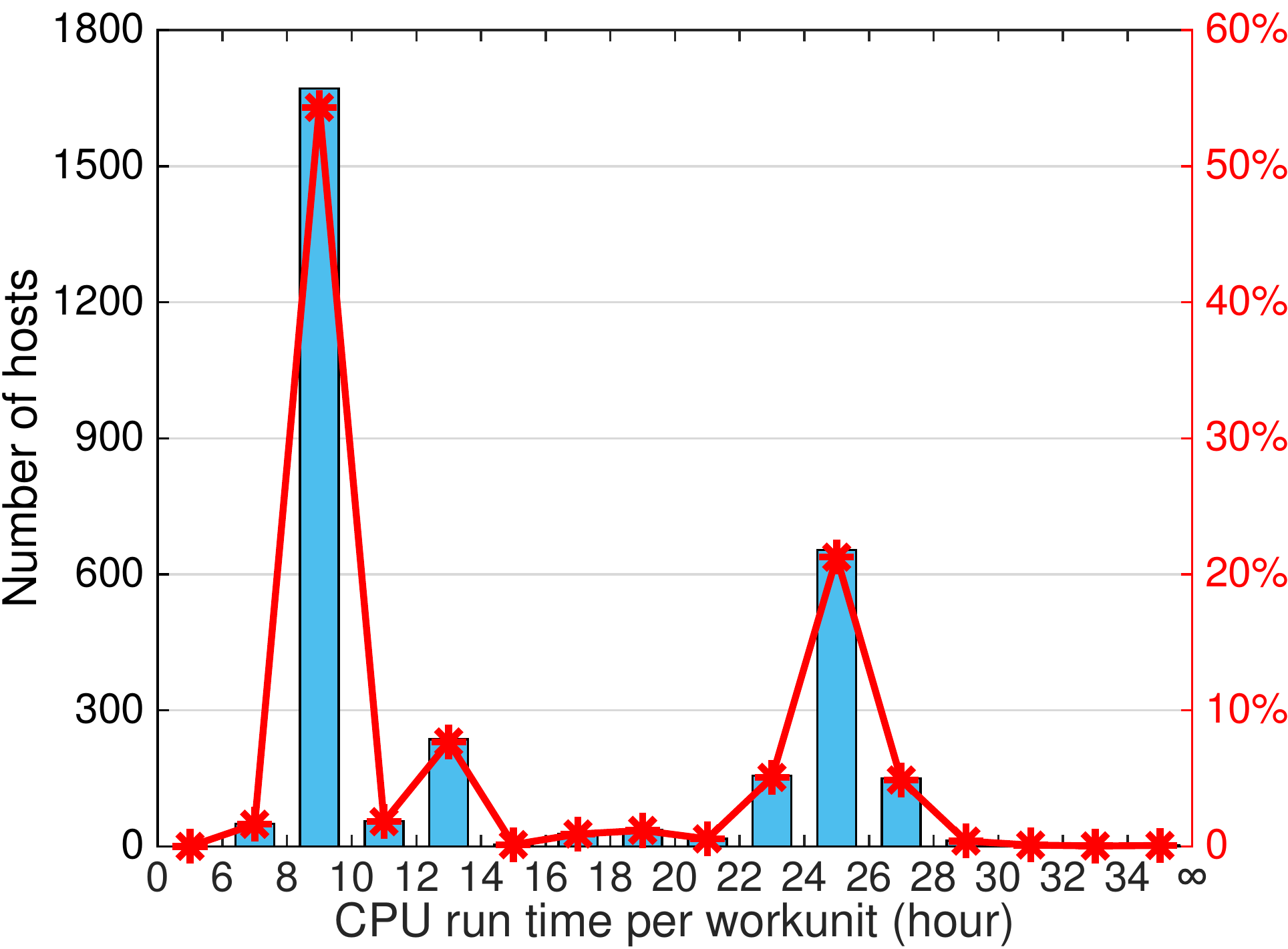}
\caption{Distribution of the run-times of the Einstein@Home O1 low frequency all-sky search work units. 
The data in this figure is taken from a sample of 3076 hosts out of total 28764 hosts of Einstein@Home. }
\label{fig:EaHtiming}
\end{centering}
\end{figure}

The computing time, and consequently the values of the timing
coefficients, in general depend on the CPU on which the search is
performed. Since the volunteer computing project Einstein@Home
comprises a broad range of different CPUs, as we optimize this search
for running on Einstein@Home for a predetermined length of time, we
need to determine how much computing power that corresponds to. A
timing analysis based on duration of the work units (WUs) of the O1
all-sky low frequency Einstein@Home search \cite{O1AS20-100}, yields
the results shown in Fig.~\ref{fig:EaHtiming}: run times are
bi-modally distributed, with a mode centered at {8-10} hours, the other at
{24-26} hours.
Based on this we divide the host population of Einstein@Home into two
types: hosts that showed a runtime of less than 14 hours were put in
one category (A) and hosts that needed more time than that were placed
in the other category (B).  The CPU models in host class (A) have an
average 8-hours runtime which is equivalent to the fast nodes on the
ATLAS computational cluster at the Albert Einstein Institute in
Hannover \cite{Atlasweb}. Therefore
the timings from fast nodes on ATLAS can be ideally used for these
Einstein@Home (A)-cores. Based on the O1 low frequency all-sky search
we estimate there are about 5300 cores in category (A) and a similar
number of cores in category (B). In the following we will assume using
only the (A)-type hosts of Einstein@Home, and hence that
one Einstein@Home month (EM) corresponds to 5300 (A)-type
cores used continuously for 1 month.
%We received  $\sim$16000 8-hours tasks averagely per day from those fast hosts in the science run of O1 all-sky search, so the the c\cite{OptimalMethod}uting power  of Einstein@Home if only counting fast hosts is $16000\times8/24=5333$ cores.
%In order to use the timings we obtained for  fast nodes of ATLAS, we assume we are going to use the fast host of Einstein@Home in this search. In the following of this paper, EM corresponds to 5333 fast CPU cores round the clock for one month.

\section{Astrophysical priors}
\label{subsec:astroprior}

An important feature of our optimization scheme is that it forces us 
to explicitly incorporate our astrophysical priors on the signal parameters.  In particular, we need to choose
the astrophysical priors on the following parameters for each target:
\begin{itemize}
\item the age and distance of the target sources. The age influences
  the spindown range that should in principle be searched. The
  distance of the source directly influences the amplitude $h_0$ of
  the signal. For some targets such as Cas A there is very little
  uncertainty in the distance and/or age of the object, 
  so the prior is chosen as a delta function. Other
  sources like Vela Jr., for instance, have large uncertainty in age
  and distance. However as was done in \cite{OptimalMethod}, we pick
  delta-functions at the extremes of the possible range. In doing this we might also include non-physically motivated age-distance combinations. We do this only to give a sense of the impact of different priors on the final results. 
  \item The star's ellipticity $\varepsilon$ and the fraction $x$ of
  spin-down energy carried away in GWs influence the GW signal
  amplitude $h_0$. For the former we pick a log-uniform distribution;
  for the latter we pick a value inspired by the upper limits
  measured with targeted searches \cite{knownPulsarsO1}
\item The $P_c$ depends not only on $h_0$, but also on the prior
  probability $P(f,\dot{f},\ddot{f})$ that the actual signal is in the
  cell defined by specific values of $(f,\dot{f},\ddot{f})$.
\end{itemize}

The results in \cite{OptimalMethod} indicate that the distance of an
object is the most important parameter in determining its GW
detectability. Among the targets considered in \cite{OptimalMethod}
targets Vela Jr. is the only source which has large uncertainties both
in age and distance.  However, even assuming 
a fairly pessimistic value of 750 pc for the distance of Vela Jr. \cite{VelaJrDist2015}, it still contributes the most to the total detection probability with respect to all other
targets.  We pick the four extremes for the priors on age and distance
of Vela Jr.: close and young (CY), close and old (CO), far and young
(FY) and far and old (FO) Vela Jr..  It is important to keep in mind
that the astrophysically viable alternatives are CY and FO and there
is no support for CO and FY {\footnote{{To estimate the age for Vela Jr., one measures its angular size $\theta$ and expansion rate $R$.  It follows that the ratio of the distance $D$ and the age $\tau$ is given by $D/\tau = R/\theta$.  Thus one does not measure $D$ or $\tau$ independently but only the ratio and so a larger $D$ implies a larger $\tau$.  Therefore we have astrophysical support for CY and FO and not for CO or FY. }}}. However, we shall include all four
alternatives for the purpose of illustrating the impact on the
optimization scheme.  Realistic searches will obviously consider only CY and FO.

Table~\ref{tab:sources} details the parameters of the three objects
that we consider in this paper.  Searches for signals from one of the
youngest known SNR, Cas A, have been carried out with LIGO data
\cite{S5CasA, S6CasA,S6NineYoungSNRs}. The reason for targeting this
source is that a young object is more likely to be spinning down
faster and hence there is more kinetic energy that could potentially
be radiated away in GWs.  Here we will see that Cas A is the third
source that contributes the overall detection probability.  Vela
Jr. and G347.3 contribute much more to the total detection probability
and are thus more promising targets.  While \cite{S6NineYoungSNRs}
presents searches and upper limits for Vela Jr. and G347.3, these have
not been the primary targets for any CW search to date. Thus there
have not been any deep CW searches for these two objects so far.
\begin{table*}
\caption{
  \label{tab:sources}
  Point source targets considered in this paper }
  \begin{center}
  \begin{tabular}{cccccccc}
   % \begin{tabular}{llllllll}
    \tableline
   
    SNR G name & Other name & Point source~J & $D_\mathrm{kpc}$ & 
    $\tau_\mathrm{kyr}$ 
    \\
    \tableline
    111.7\textminus2.1 & Cas~A & 232327.9+584842 & 3.3--3.7 & 0.31--0.35
    \\
    266.2\textminus1.2 & Vela~Jr. & 085201.4\textminus461753 &  0.2--0.75 & 0.7--4.3
    \\
    347.3\textminus0.5 & & 171328.3\textminus394953 &  1.3 & 1.6
    \\
    \tableline
  \end{tabular}
\end{center}
\end{table*}

The ellipticity $\varepsilon$ is the least known parameter so here we take a flat probability density on $\log\varepsilon$ within a conservative range of values. 
We target weak signals and hence the maximum value of  $\varepsilon$ we allow in the $i$th cell is:
\begin{equation}
  \label{eq:epsilonMaxAge}
    \varepsilon^\mathrm{max}_i = \min ( {~\varepsilon^\mathrm{sd}_i,~\varepsilon^\mathrm{age}_i,~10^{-6}} )\,,
\end{equation}
where $\varepsilon^\mathrm{sd}_i$ and $\varepsilon^\mathrm{age}_i$ are the spin-down ellipticity upper limit and the spindown age-based ellipticity upper limit respectively: 
\begin{equation}
  \label{eq:spindownEllipticityX}
  {\varepsilon^\mathrm{sd}}=\sqrt{ \frac{5c^5}{32\pi^4G}\frac{x|\dot{f}|}{If^5}}\,.
\end{equation}
and
\begin{equation}
  \label{eq:eliimitage}
  {\varepsilon^\mathrm{age}}= \frac{c^2}{16\pi^2f^2}\sqrt{ \frac{10c}{GI\tau}}\,.
\end{equation}
In Eqs.~(\ref{eq:spindownEllipticityX}) and (\ref{eq:eliimitage}), $f$ is the instantaneous
frequency of the emitted GW signal, $G$ Newton's constant, $c$ the speed of light, and $\tau$ the age of the source. $I$ is the the moment of inertia and we use its standard
value $10^{38}\,\mathrm{kg \,m^2}$ for all the results in this paper.
$x$ is the fraction of  spin-down energy loss due to GW emission.
The latest observational limits on GW emission from the Crab and Vela pulsars
constrain $x$ to less than $0.2\%$ and $1\%$ respectively \cite{knownPulsarsO1}. In this paper, we will assume $x=1\%$. There is no guarantee that the Vela and Crab results apply to these other objects, but right now they are the only measurements at hand.
According to the results of \cite{Horowitz}, the realistic maximum value of $\varepsilon$ is expected to be
smaller than $4\times10^{-6}$. Hence the third limit we use in this paper is $10^{-6}$.
We take $\varepsilon^\mathrm{min}= 10^{-14}$ because deformations of a compact star due to the internal magnetic field (at least $10^{11}$ G)  are not expected to be smaller than  $\sim10^{-14}$  \cite{Andersson:2009yt}. 
Based on the above discussion, our prior $p(\varepsilon)$ is: 
\begin{equation}
  \label{eq:epsilonPrior}
  p(\varepsilon) = \left\{ 
    \begin{array}{cc}
     {1\over\varepsilon} {1\over{\log(\varepsilon^\mathrm{max}/\varepsilon^\mathrm{min})}}\, &\quad  \varepsilon^\mathrm{min}<\varepsilon<\varepsilon^\mathrm{max}   \, \\
      0\,   & \mathrm{elsewhere} \,.
    \end{array}   
  \right. 
\end{equation}

Since the GW frequencies emission frequencies are unknown, our search encompasses a large range; namely, from 20 to 1500 Hz. For a given $f$, the $\dot{f}$ and $\ddot{f}$ ranges are determined by the fiducial age of the source $\tau$:
\begin{equation}
\label{eq:Priors}
	\begin{cases}
	20\,\mathrm{Hz} \le  f ~\le 1500\,\mathrm{Hz}\\
	-f/ (n-1) ~\tau\, \le   \dot{f} ~\le 0\,\mathrm{Hz/s}\\
	0\,\mathrm{Hz/s}^2 \leq  \ddot{f} \leq ~n{f}^/\tau^2.
	\end{cases}
\end{equation}
{$n$ is the braking index. If the frequency evolution follows $\dot{f}\propto f^{n}$,  the second order spindown is then  $\ddot{f}=n\dot{f}^{2}/f$.}
In the second equation we take $n=2$ to encompass the broadest range of $\dot{f}$ values. In the third equation we assume a braking index $n=5$, corresponding to phase evolution
purely due to GW emission, and a constant $\dot{f}$ value of $-f/\tau$. For all other mechanisms $n<5$ and in
particular for pure dipole electromagnetic emission $n=3$ (see
e.g. \cite{ShapiroTeukolsky}). Therefore, our search ranges for $\dot{f}$ and
$\ddot{f}$ in (\ref{eq:Priors}) encompass all combinations of
emission mechanisms. Since none of the quantities that determine the detection probability at a given frequency, such as the maximum ellipticity or the noise level, depend on $\ddot{f}$ we drop the $\ddot{f}$ dependence in the
signal probability density $P(f,\dot{f})$, equivalent to a uniform prior on the $\ddot{f}$ range. We consider both uniform
and log-uniform priors on the search range of $f$ and $\dot{f}$ reflecting our ignorance on those signal
parameters.

In this search, we do not search over the third order spin-down $\dddot{f}$ because, even 
for the youngest target, a search over third order spin downs is not necessary for the coherent and observation times that we are considering here.

\section{The optimization}
\label{sec:OpSetup}

The optimization scheme is introduced in \cite{OptimalMethod}.  
%Wecassume that we can estimate the computing cost and detection probability for any given parameter space volume and search set-up.
The starting point is to divide the parameter space into
non-overlapping small cells such that the computing cost to search each cell and the resulting detection probability are roughly constant within each cell.  The cost for each
should also be much smaller than the full computing cost budget
available to us.  For each cell, we calculate the computing cost and
detection probability. For each cell we define the
efficiency, that is the ratio of detection probability to computing
cost.  In the absence of any constraint apart from the total budget,
one could proceed simply by picking the most efficient cells till the
computing budget is exhausted.  However, we do have an additional
constraint, namely that we do not want to search the same parameter
space cell with multiple search set-ups.  It is shown in \cite{OptimalMethod}
how this constraint can be included in the optimization using linear
programming techniques.

The search set-ups are defined by 6 parameters: the
segment coherent duration $T_{\mathrm{coh}}$, the nominal mismatch
parameters $m_f$, $m_{\dot{f}}$, $m_{\ddot{f}}$, and the refinement
factors $\gamma^{(1)}$ and $\gamma^{(2)}$ (see details in
Section~\ref{sec:templateBanks}).  For each given search set-up we can
compute the mismatch distribution, i.e. the distribution of the
fractional loss in signal-to-noise ratio due to the discreteness of
the template bank\footnote{Mismatch distributions of these set-ups are
  computed for Vela Jr., since it is responsible for most of the
  detection probability}.  

The optimization not only determines the best search set-ups but also
how to distribute the computing budget among the different
astrophysical targets and over the parameter space.  In
\cite{OptimalMethod} we presented this optimization scheme under a set
of simplifications. We now take \cite{OptimalMethod} as the starting
point for an actual search design but here we forego the
simplifications that we made in \cite{OptimalMethod} and take into
account issues of feasibility and practicality.  We shall take
rectangular cells in $(f,\dot{f})$ space, and each cell will be
$10\,$Hz wide in frequency and $10^{-9}\,$Hz/s in $\dot{f}$.  To
completely specify the parameter space cells, we need to include
$\ddot{f}$, and each of the $(f,\dot{f})$ cells are allowed to take
all permissible values of $\ddot{f}$.  As mentioned above, for each
cell, each search set up, and every choice of astrophysical prior, we
calculate the computing cost and detection probability.  Since it is
not possible to predict the mismatch distribution of the GCT search
for a given set-up, these are produced from thousands of
injection-and-recovery Monte Carlos.

In principle, the above ingredients are sufficient for the
optimization scheme.  However, certain obvious simplifications can be
made based on practical considerations:
\begin{itemize}

\item Among all the set-ups with the same $T_{\mathrm{coh}}$, for
  every value of the computing cost, we select the set-up which has
  the lowest average mismatch.  In our case, as we shall describe in
  greater detail below, this results in 71 seeded set-ups for all the
  different values of $T_{\mathrm{coh}}$ considered.

\item Among all the cells, each with 71 set-ups, we select those so
  that the sum of relative detection probabilities constrained by the
  available computing budget, is maximized. This is done through
  linear programming.

\item Since we find that the optimal choice of cells spans a very
  broad frequency range, we determine the loss in detection
  probability caused by extending the search parameter space to
  include the broadest frequency band. We further evaluate the loss
  incurred with respect to the optimal solution due to
  utilizing the same set-up across all cells for every
  astrophysical target. Both these choices, when viable, greatly
  simplify the post-processing of the results.

\item We compare the final results from different sets of priors,
  estimate the loss in detection efficiency due to having optimized
  assuming a wrong age and choose the set-up such that this loss is
  the smallest.
\end{itemize}

In the next sections we describe the above steps in greater
detail.
% \begin{figure}
% \begin{centering}
% \includegraphics[width=0.50\textwidth]{optimization_scheme.pdf}
% \caption{Optimization scheme flow chart }
%  \label{flow_chart}
% \end{centering}
% \end{figure}

%??????????????????????

\subsection{Set-ups: primary selection}
\label{subsec:PS}

Running the optimization scheme on $\sim$ 2000 set-ups is
computationally too burdensome. We hence down-select, among the ones
with the same $T_{\mathrm{coh}}$, those that yield the lowest average
mismatch at fixed computing cost over the entire prior parameter range
of Vela Jr.. In principle this selection should be done separately for
every target, but here we simplify the procedure in this manner
because Vela Jr. contributes the bulk of the detection probability.

\begin{figure}
\centering
\includegraphics[width=0.45\textwidth]{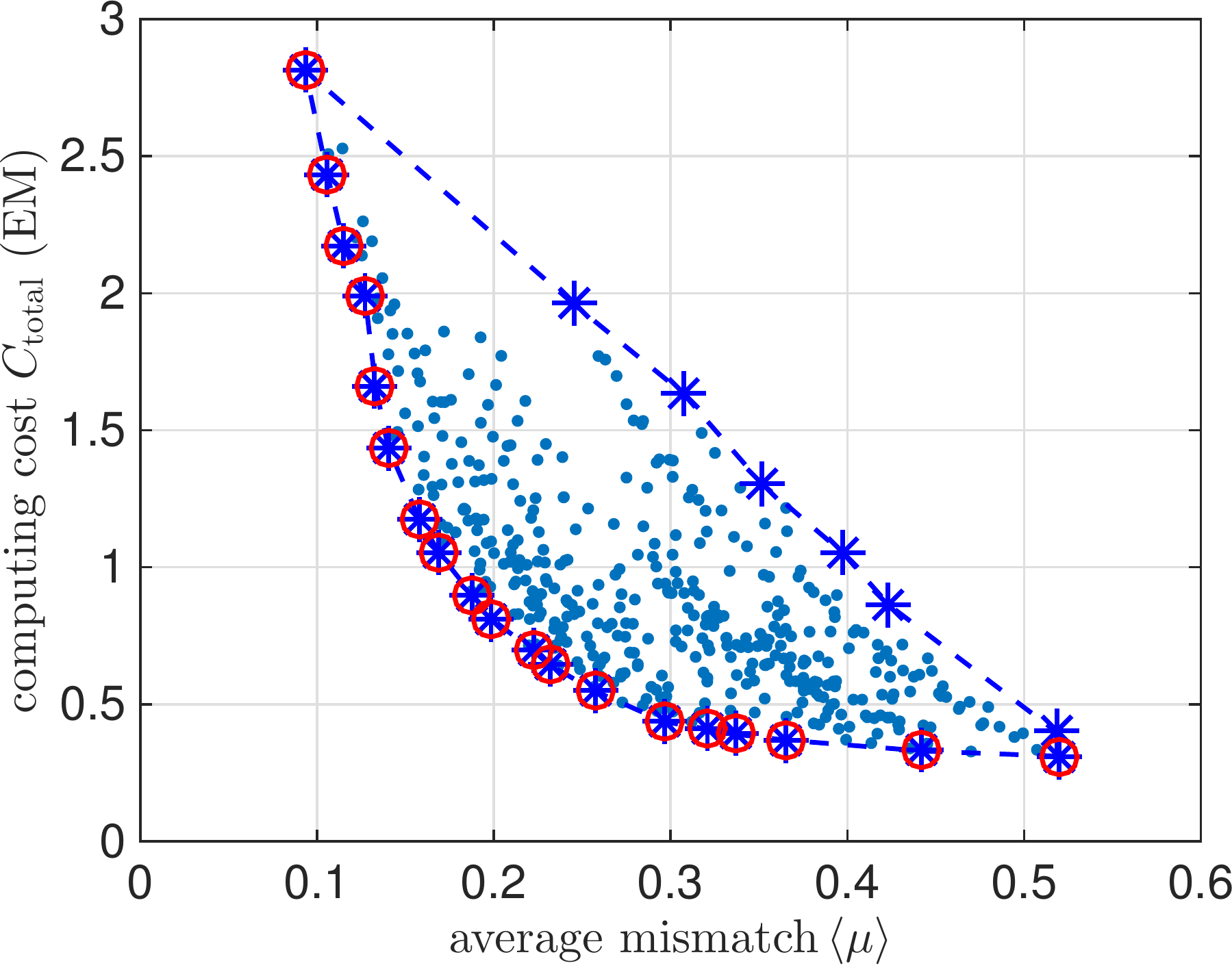}
\caption{432 set-ups with $T_{\mathrm{coh}}=15$ days. The 19 blue stars in red circles are selected for further consideration as explained in the text.}
 \label{scatter_15D}
\end{figure}

Fig.~\ref{scatter_15D} shows computing cost and average mismatch for 432 set-ups corresponding to different grid spacings and $T_\mathrm{coh}=15$ days. Out of these, 19 are selected which have the lowest  measured average mismatch at fixed computing cost. Considering different values for $T_\mathrm{coh}$ we select 71 set-ups : 13 are from $T_\mathrm{coh}=10$ days, 19 from $T_\mathrm{coh}=15$ days, 15 from $T_\mathrm{coh}=20$ days, 14 from $T_\mathrm{coh}=30$ days and 10 from $T_\mathrm{coh}=60$ days. Since the core of the optimisation procedure is linear programming and its computing time grows at least polynomially with the number of tested set-ups \cite{TRAUB198259}, it is easy to see that going from 2000 set-ups to 71 reduces the cost by at least a factor of $>$ 100. For reference we note that it took $\sim$10 CPU hours to perform the optimisation with the 71 set-ups.

%\mapcomment{check the following statement:} 
The {\it measured} average mismatch $\langle \mu \rangle$ of each of these distributions is used to reduce the expected signal-to-noise ratio $\rho^2$ of a putative signal. More specifically, with respect to Eq.~(16) of \cite{OptimalMethod} that used only $\rho^2$, we now use a more realistic estimate of the actual results of a search:
\begin{equation}
\label{eq:newSNR}
\rho^2 \rightarrow (1- \langle\mu\rangle) \times \rho^2.
\end{equation}
This estimate folds-in the sensitivity-loss effect of using finite grids and does it realistically because it is based on the measured mismatch distributions of the actual search codes.
%In Fig.~\ref{scatter_15D}, 432 Set-ups (dots) with $T_\mathrm{coh}=15$D are scattered on $\langle m\rangle-C_\mathrm{total}$ plane. Here the $C_\mathrm{total}$ is c\cite{OptimalMethod}uted for the full search parameter space of Vela Jr (young case). 25 interesting set-ups (blue stars) show up on the edge of  the shape formed by dots. If one star has higher  $\langle m\rangle$ and $C_\mathrm{total}$ than any other star's, it is excluded. Hence, 19 blue stars in red circle are selected as seeded set-ups.  Because the detection probability depends not only on $\langle m\rangle$ on but also on $T_\mathrm{coh}$, we have processed this  primary selection for every $T_\mathrm{coh}$. Finally, 71 in total set-ups are selected as seeded set-ups among which 13 are from $T_\mathrm{coh}=10$D, 19 from $T_\mathrm{coh}=15$D, 15 from $T_\mathrm{coh}=20$D, 14 from $T_\mathrm{coh}=30$D and 10 from $T_\mathrm{coh}=60$D.

\subsection{Optimization under different assumptions}
\label{subsec:rep}

Following Section~V of \cite{OptimalMethod}, we now use linear programming to determine how to best pick targets, waveform parameter space to search and search set-ups under a set of different assumptions on the age and distance of the source. We consider 71 different set-ups. A significant difference with respect to \cite{OptimalMethod} is that here we consider a number of different grid spacings for the same $T_{\mathrm{coh}}$ and, as explained above, {\it the actual} mismatch associated with each of them. 

\begin{figure*}%
    \centering
    \subfloat[Assuming Vela Jr. is close and young]{{  \includegraphics[width=.45\linewidth]{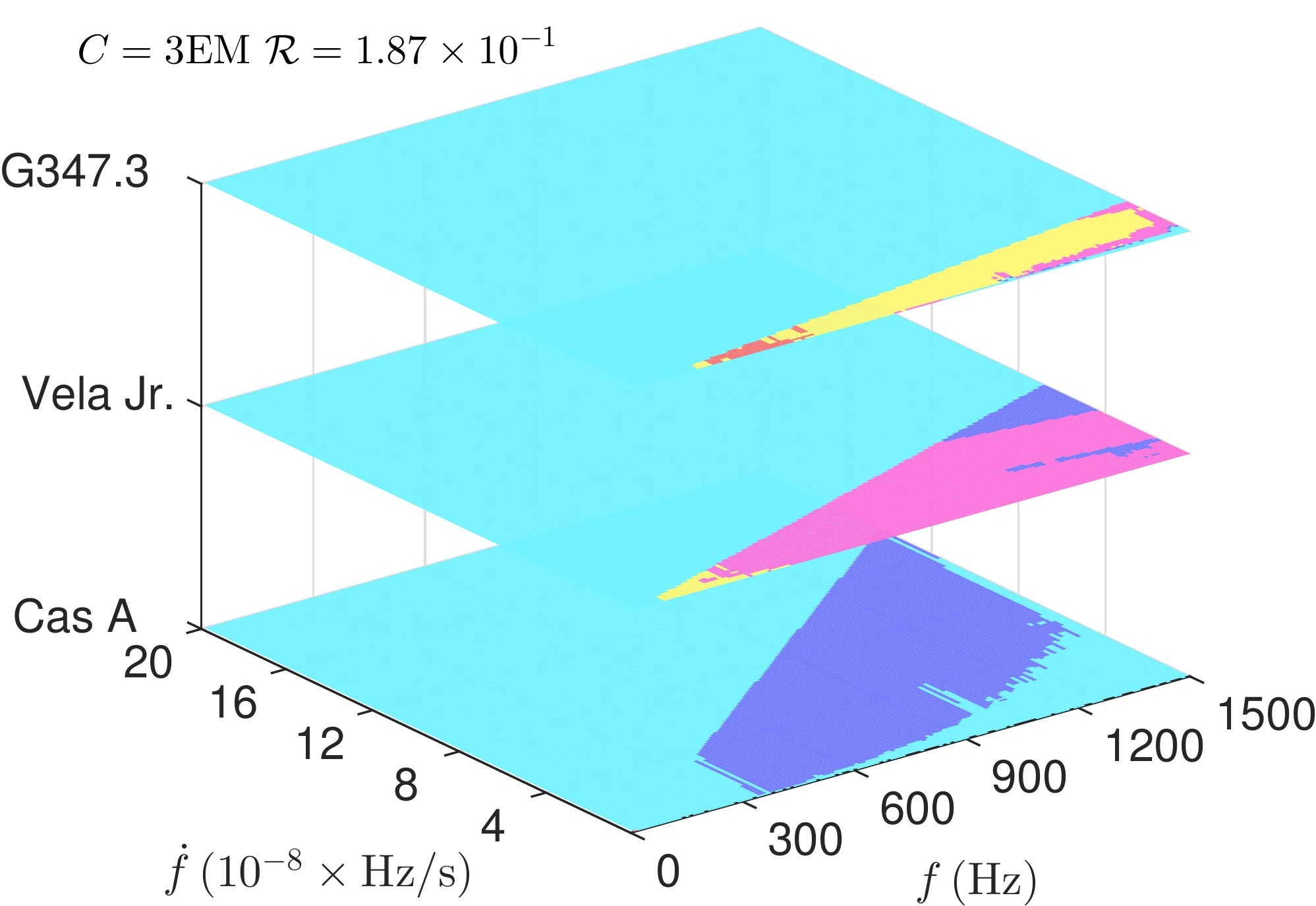}}}%
    \qquad
    \subfloat[Assuming Vela Jr. is close and old]{{  \includegraphics[width=.45\linewidth]{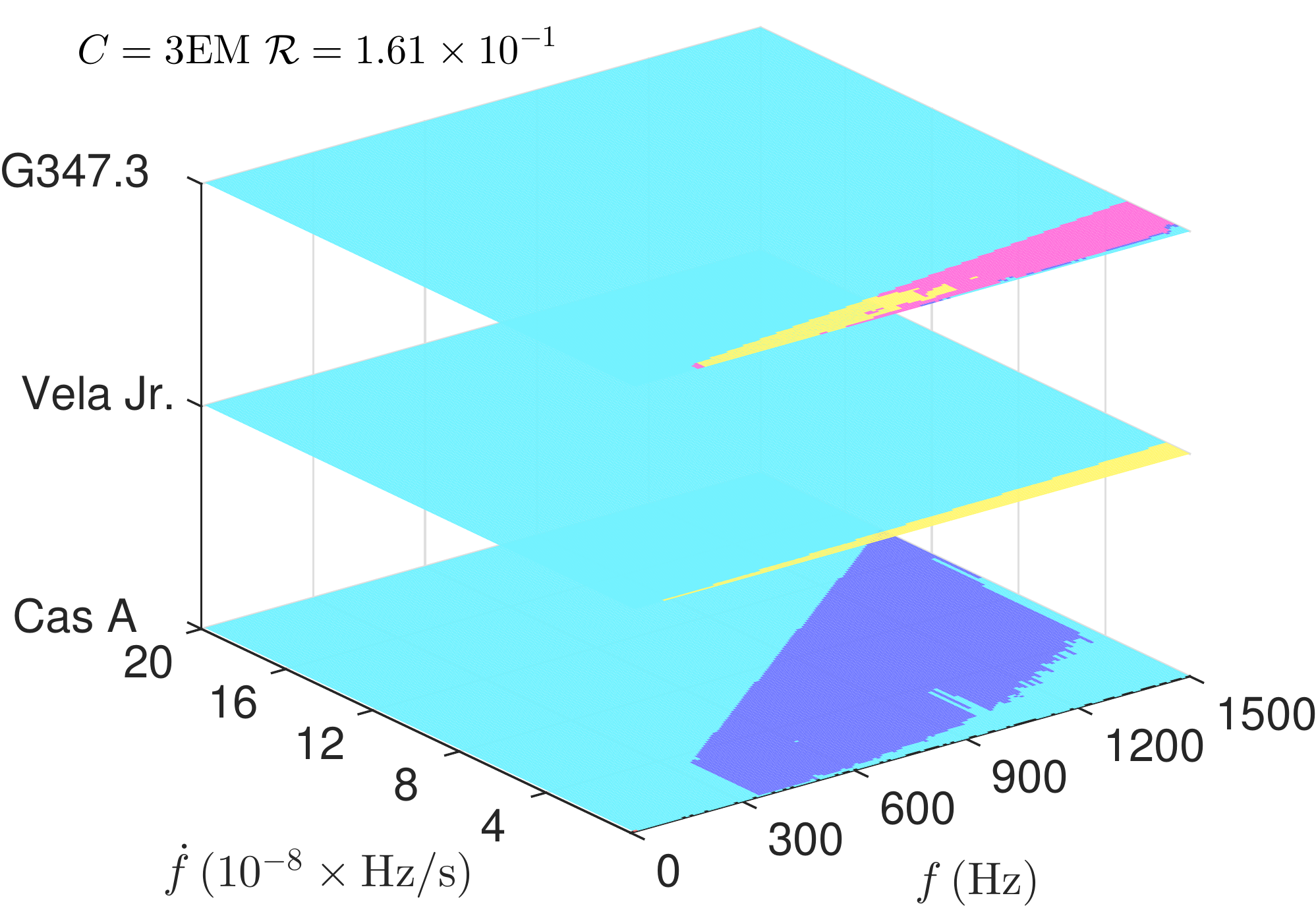}}}%
   \qquad
    \subfloat[Set-up details on the plane of Vela Jr. (assuming Vela Jr. is close and young)]{{  \includegraphics[width=1\linewidth]{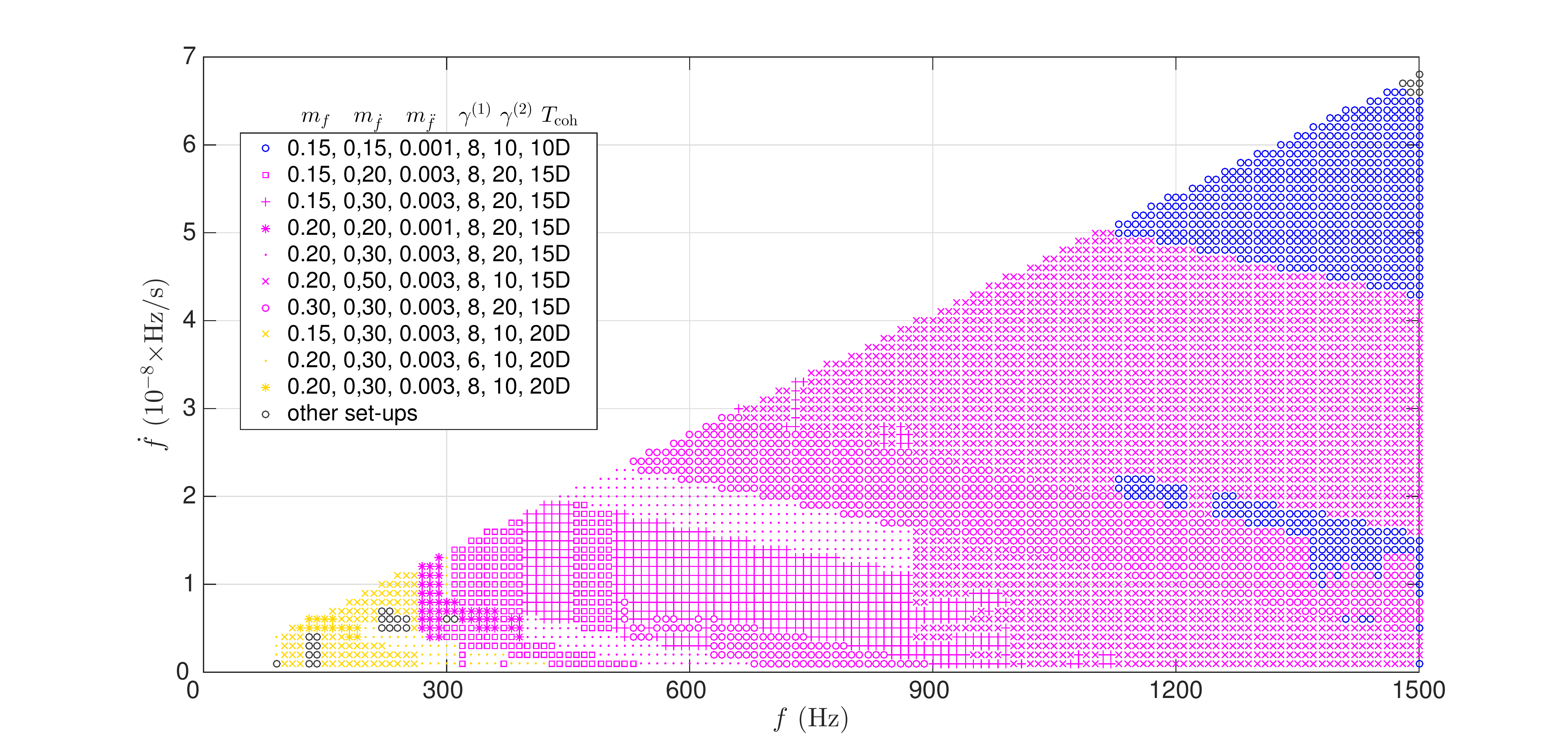}}}%
    \caption{Parameter space coverage assuming uniform $f,\dot{f}$ priors, Vela Jr. at 200 pc (C) and 3 EM computing budget. In panel (a) we assume Vela Jr.'s age is 700 years  (Y) and in panel (b) we assume that Vela Jr. is  4300 years old (O). The plane is indicated in aqua. Cells in blue are searched with 10-day $T_\mathrm{coh}$ set-ups, magenta indicates 15-day $T_\mathrm{coh}$, yellow the 20-day $T_\mathrm{coh}$, red the 30-day $T_\mathrm{coh}$, and green the 60-day $T_\mathrm{coh}$ (although not used in either  cases). 
The computing power used on Cas A, Vela Jr. and G347.3 is 1.36 EM, 1.14 EM and 0.50 EM, respectively, for the set-ups of panel (a), and 1.97 EM, 0.21 EM and 0.82 EM for the set-ups of panel (b). The contribution to the total detection probability from Cas A, Vela Jr. and G347.3 is 1.3\%, 14.4\% and 3.1\% for panel (a), and 1.4\%, 11.5\% and 3.1\% for panel (b).  
Note that each color represents set-ups which have the same coherent duration, but might differ in the grid spacings. Figure (c) shows the set-up details on Vela Jr. plane in Figure (a). For example, the blue circles represent the cells that need to be searched by using the set-up: $m_f=0.15$, $m_{\dot{f}}=0.15$, $m_{\ddot{f}}=0.001$, $\gamma^{(1)}=8$, $\gamma^{(2)}=10$, and $T_\mathrm{coh}=10$ day.}%
 \label{fig:cyco}%
 \end{figure*}

Fig.~\ref{fig:cyco} shows the results of the optimization under the assumptions that Vela Jr. is at a distance of 200 pc (C), that the signal frequency $f$ and spindown $\dot{f}$ are uniformly distributed within their ranges (defined in Eq.~(\ref{eq:Priors})) and with a 3 EM computing budget.

   \begin{figure*}%
    \centering
    \subfloat[G347.3 (close and young for Vela Jr.)]{{  \includegraphics[width=.45\linewidth]{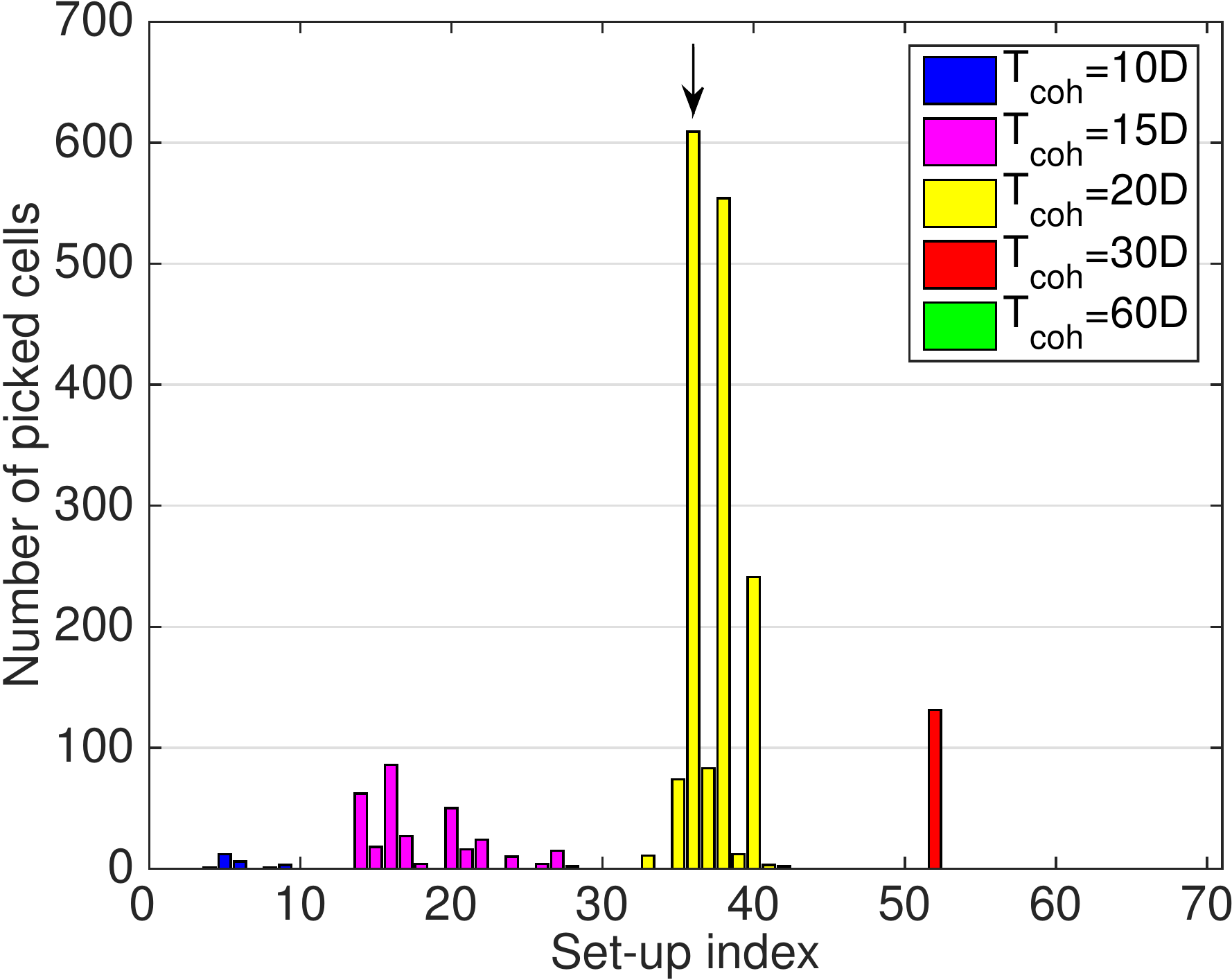}}}%
    \qquad
    \subfloat[G347.3 (close and old for Vela Jr.)]{{  \includegraphics[width=.45\linewidth]{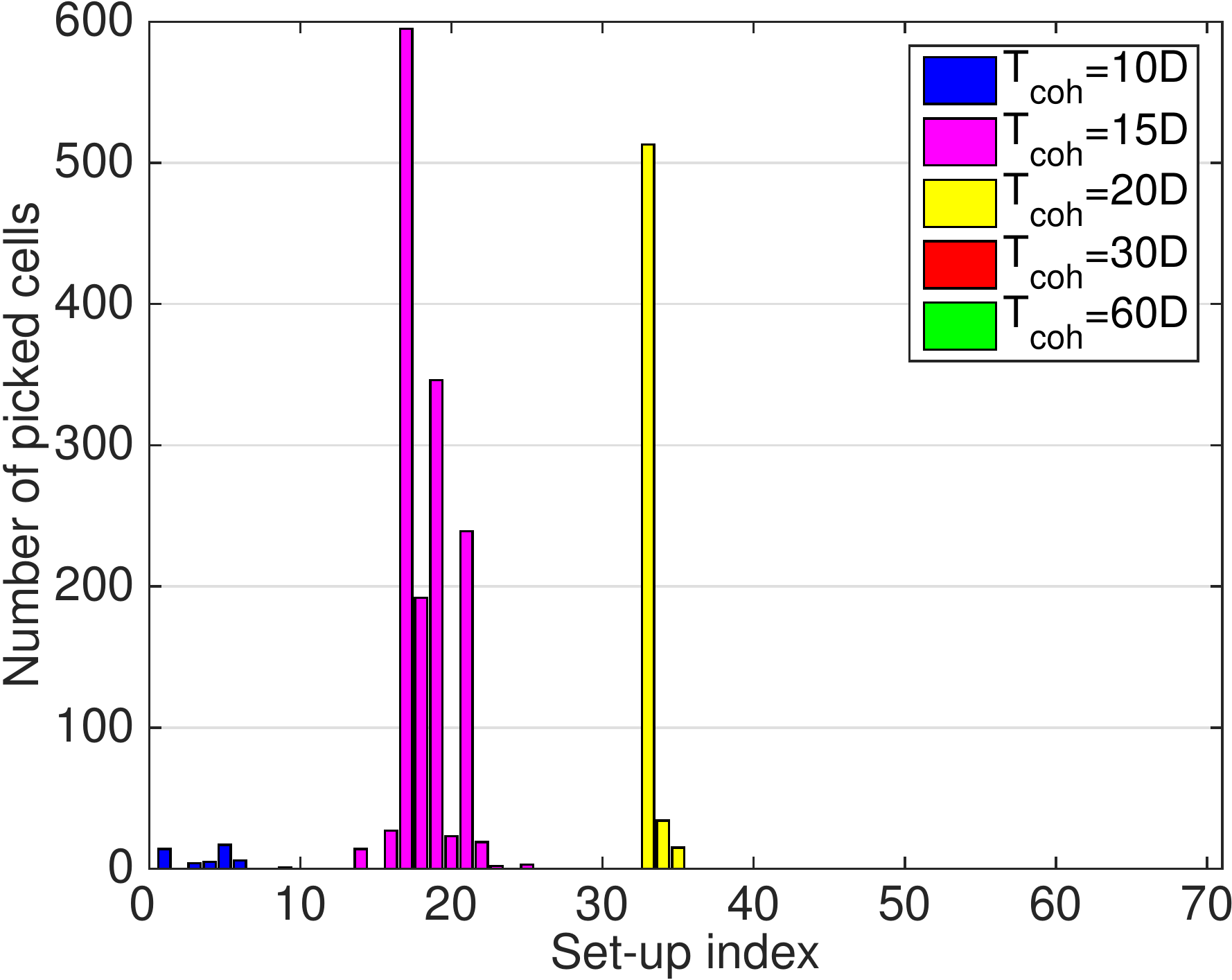}}}%
    \qquad
    \subfloat[Vela Jr. (close and young for Vela Jr.)]{{  \includegraphics[width=.45\linewidth]{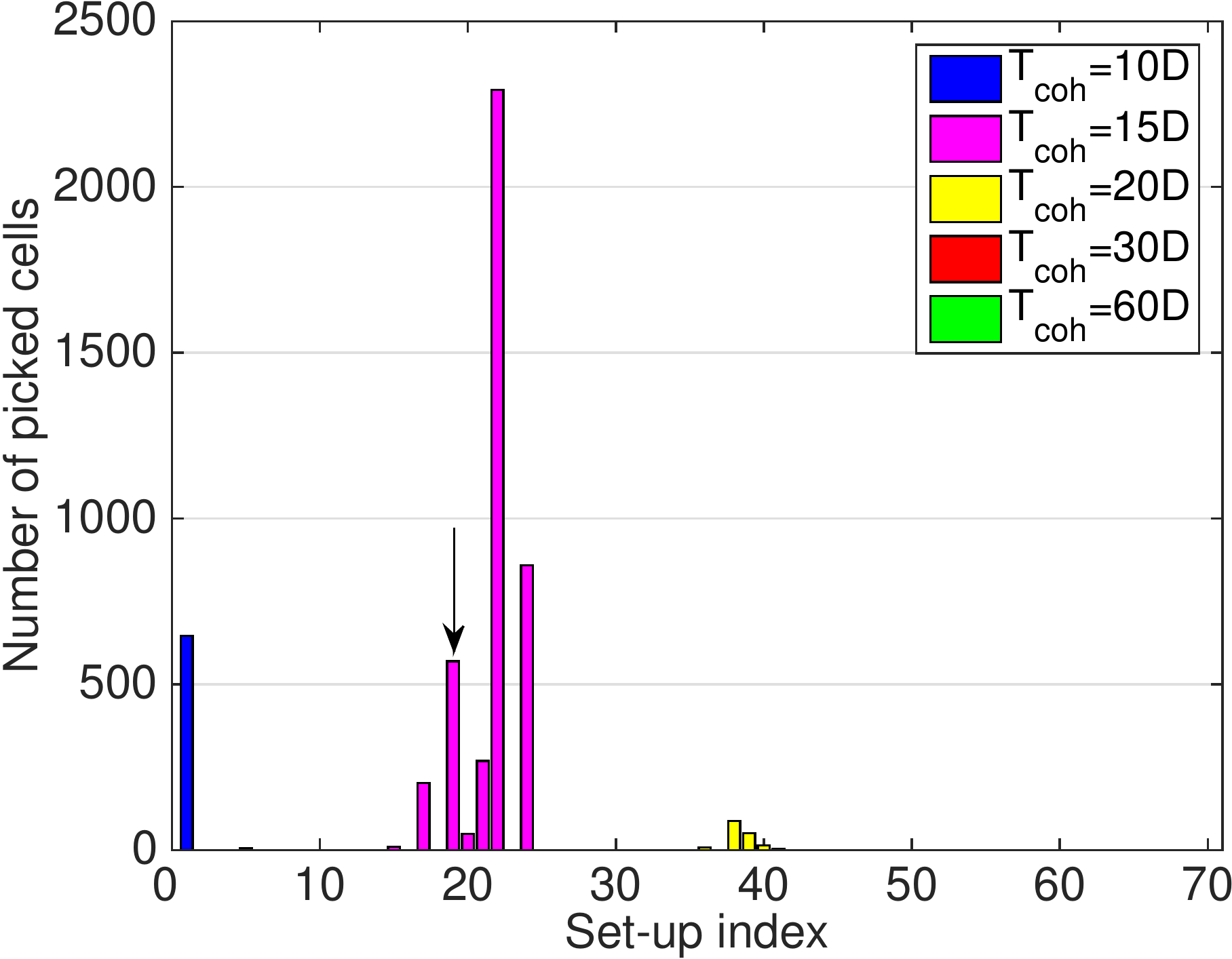}}}%
    \qquad
    \subfloat[Vela Jr. (close and old for Vela Jr.)]{{  \includegraphics[width=.45\linewidth]{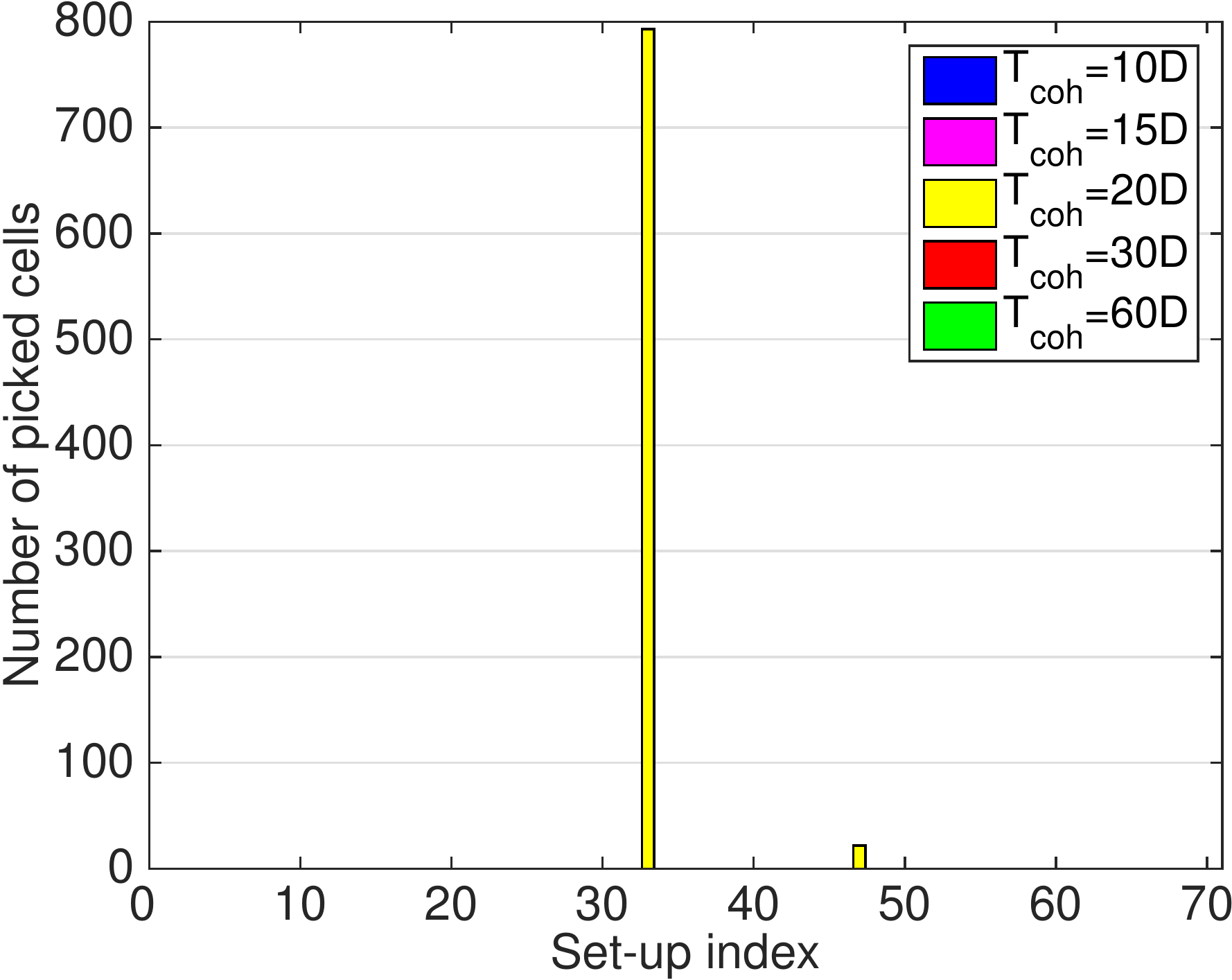}}}%
    \qquad
    \subfloat[Cas A (close and young for Vela Jr.)]{{  \includegraphics[width=.45\linewidth]{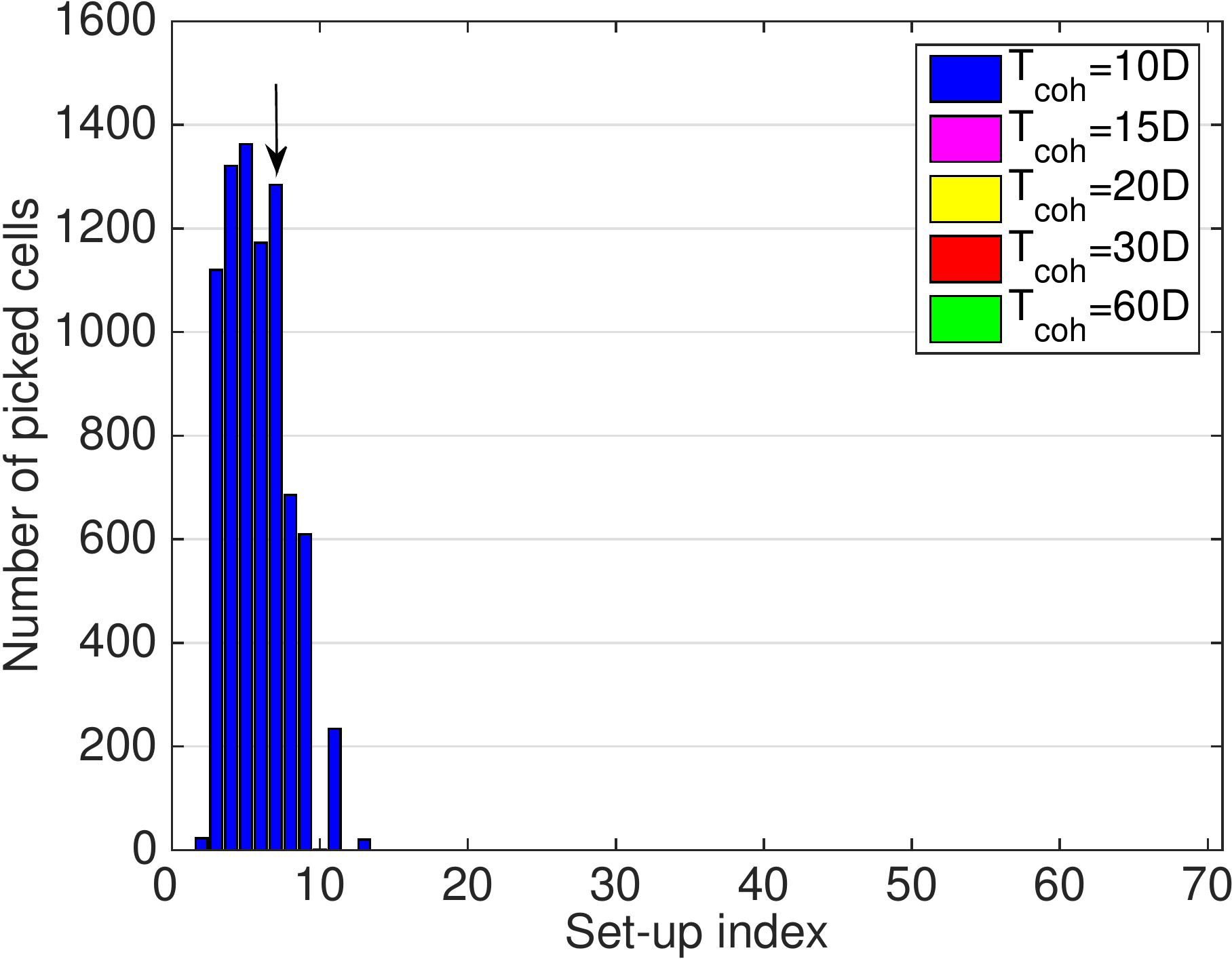}}}%
    \qquad
    \subfloat[Cas A (close and old for Vela Jr.)]{{  \includegraphics[width=.45\linewidth]{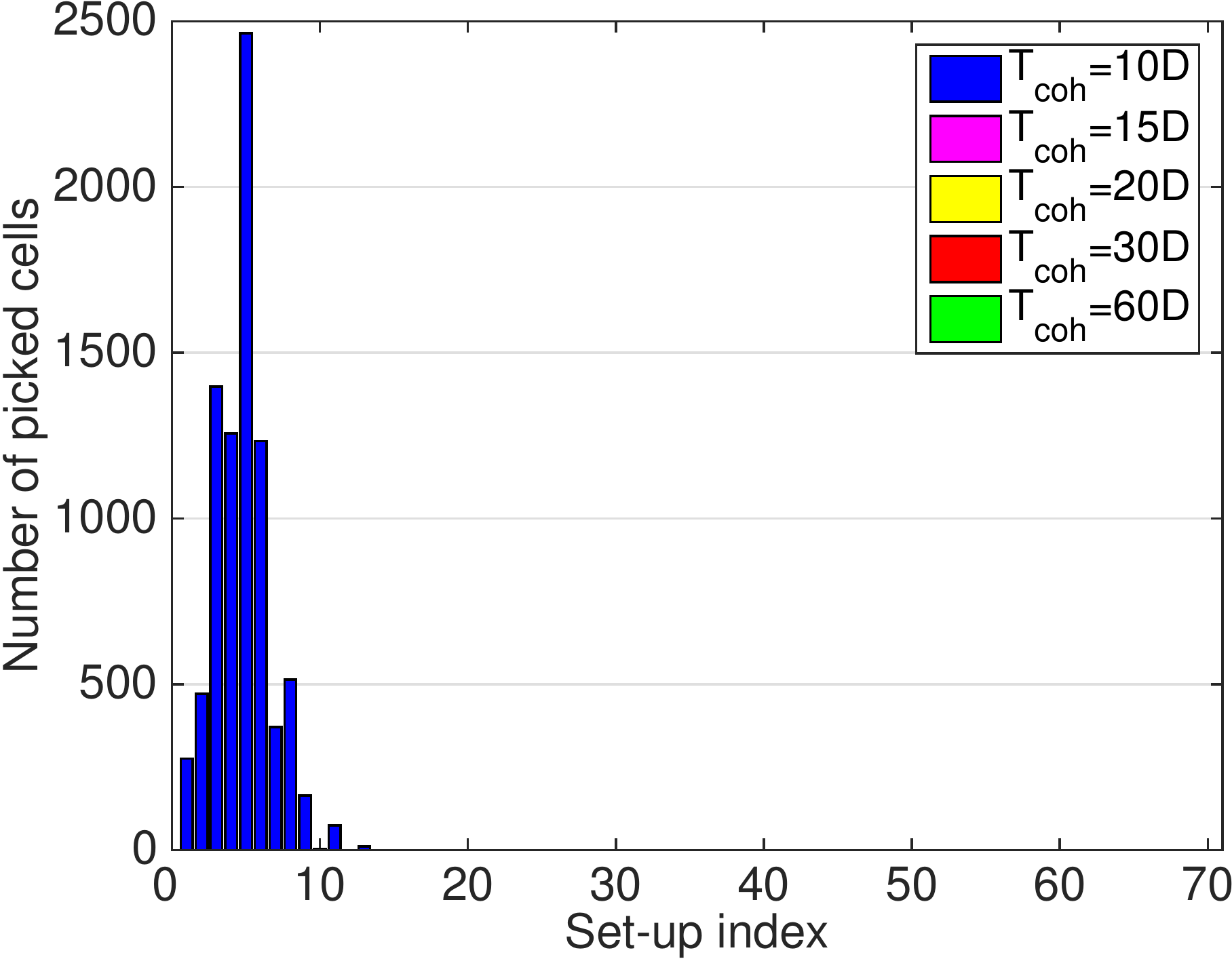}}}%
    \caption{Optimal set-up details. The left hand-side figures show the set-ups assuming Vela Jr. is close and young.  The right hand-side figures show the set-ups assuming Vela Jr. is close and old. As explained in in Section \ref{subsec:simply}, the arrows indicate the set-ups that will finally be adopted in the search.}%
 \label{spDetail}%
\end{figure*}

Fig.~\ref{fig:cyco} shows the result of the optimization
procedure. Different colors represent search set-ups having different
$T_\mathrm{coh}$. The total detection probability $\mathcal{R}$ is
defined in Eq.~(49) of \cite{OptimalMethod} as the sum of detection
probabilities of the selected parameter space cells.  $C$ indicates the total computational cost and
in general this will be equal to or smaller than the maximum budgeted
3EM. The results for Vela Jr. for the CY and CO cases can be summarized as follows: 
\begin{itemize}
\item When Vela Jr.'s age is assumed to be 700 years, the spin-down
  parameter space to search is large and the $T_\mathrm{coh}$ chosen
  set-ups for Vela Jr. are 10 days, 15 days and 20 days (see
  Fig.~\ref{fig:cyco}(a)). Fig.~\ref{fig:cyco}(c) shows the Vela Jr.
  plane of Fig.~\ref{fig:cyco}(a) and we can appreciate that more than
  10 different set-ups constitute the optimal search for young Vela
  Jr.

\item When Vela Jr.'s age is assumed at 4300 years, the spin-down
  parameter space shrinks and the $T_\mathrm{coh}$ chosen set-ups for
  Vela Jr. are all 20 days (see Fig.~\ref{fig:cyco}(b)). This is the CO
  case and we emphasize again that there is no astrophysical support
  for this scenario.  We include this to illustrate the effect of the
  priors on the optimization results.
\end{itemize}

The total detection probability $\mathcal{R}$ corresponding to the two different age priors is  also different: 18.7\% and 16.1\% for the young and for the old Vela Jr., respectively. Under either assumptions Vela Jr. contributes the bulk of the total  probability: 14.4\% out of 18.7\% and 11.5\% out of 16.1\%. However in the former case the Vela Jr. search uses up 38\% of the computing budget whereas in the latter it only uses 7\%. 
%On the top of each sub-figures, apart from $C$ and $\mathcal{R}$, the other six numbers are the computing cost in Cas A, Vela Jr, G347.3 and the detection probability from Cas A, Vela Jr, G347.3 respectively. In Fig.~\ref{fig:cyco}(b), due to the search parameter space of Vela Jr is small, the computing cost on Vela Jr is as less as 0.21 EM which is only 7\% of total budget. However, 11.5\% out of 16.1\% of detection probability are contributed from Vela Jr. Comparing these two sub-figures, you can find the chosen set-ups  and as well as the computing cost ratios are much different especially on targets Vela Jr and G347.3. This raises a very important question: which recipe of this two we are going to use in our search?

Figs.~\ref{spDetail} reveals further details of the chosen set-ups. In particular panels (a), (c) and (e) show how many cells are searched with each of the different set-ups for G347.3, Vela Jr. (CY) and Cas A, respectively. Panels (b), (d), (f) show the same quantities but under the CO assumption for Vela Jr.. Assuming Vela Jr. is 700 yr old (CY), 20-day set-ups are mostly selected for G347.3, 15-day set-ups mostly for Vela Jr., and only 
10-day set-ups for Cas A. If we assume that Vela Jr. is 4300 yr old, the parameter space of Vela Jr. shrinks due to the age limit. Only  20-day set-ups are selected for Vela Jr.. The computing savings incurred due to the smaller parameter space are re-invested in longer $T_\mathrm{coh}$. The dominant set-ups for G347.3 use shorter $T_\mathrm{coh}$ compared to the ones derived under the Vela Jr. CY prior. This is due to the fact that, with a smaller parameter space for Vela Jr., the most probability is harvested by exploiting it the most, with long (expensive) $T_\mathrm{coh}$ and this is balanced by spending less on the other targets. Finally, because of the distance of Cas A is much larger than Vela Jr. and G347.3, the optimal way to distribute the computing budget is by searching Cas A with relatively cheap (less-sensitive) 10-day set-ups.

Since we do not know the age of Vela Jr.,  the set-up corresponding to which of the two priors should we pick? To answer this question, we investigate the consequences of having picked the wrong prior, namely the impact on the detection probability if we assume that Vela Jr. is 700 yrs old when Vela Jr. is 4300 yr old and if we assume that Vela Jr. is 4300 yrs old when Vela Jr. is 700 yr old.
In the first case we search a broader spin-down range than we need to, in order to gather all the detection probability.  To do this we use a less-sensitive set-up (shorter $T_\mathrm{coh}$) for most of the frequency band for Vela Jr. and, partly also for Cas A and G347.3 (see  Fig.~\ref{spDetail}(a), (c), (e)).
The consequence of this is that we waste computing power in the high $\dot{f}$ region and lose detection probability due to the shorter $T_\mathrm{coh}$ set-ups used in the low $\dot{f}$ region. 
%Of course, another part of lost detection probability are due to the shorter $T_\mathrm{coh}$ set-ups used in parameter space of Cas A and G347.3. It is a small part because, as we discussed, Vela Jr contributes the most of the $\mathcal{R}$.
In the second case, %The second misjudgment is we overestimate the age of Vela Jr. On the Vela Jr plane, instead of using 10D,15D and 20D $T\mathrm{coh}$ set-ups in broader $\dot{f}$ region, 
we just search the low $\dot{f}$ region with longer  $T\mathrm{coh}$ set-ups (see  Fig.~\ref{spDetail}(d) ).
We gain some detection probability due to adopting longer $T\mathrm{coh}$ set-ups in the low $\dot{f}$ of Vela Jr. and in parameter space of Cas A and G347.3. Meanwhile, we lose detection probability because we give up the whole high $\dot{f}$ range of Vela Jr.. 

The results are summarized in Table~\ref{tab:mis}.
%%%%%%%%%%%%%%%table
\begin{table*}
\begin{threeparttable}
\caption{
Effect of using the wrong age prior for Vela Jr.\label{tab:mis}
}
  \begin{centering}
  \begin{tabular}[c]{c|c|c|c|c|c}
    \tableline
   
   &\multicolumn{2}{c|}{uniform distribution on $f$, $\dot{f}$~~~~}&\multicolumn{2}{c|}{log-uniform distribution on $f$, $\dot{f}$}&
   \\
   \cline{2-6}
  
   &C for Vela Jr.&F for Vela Jr.&C for Vela Jr.&F for Vela Jr.& $\langle loss \rangle$\\
   \cline{1-6}  

$\mathcal{R}$ if Vela Jr. is 4300 yrs old&
14.8\%(16.1\%):&
7.9\%(8.6\%):&
6.4\%(11.4\%):&
4.7\%(4.9\%):&\\

but prior assumes 700 yrs of age &
$loss=$\textbf{7.9\%}&
$loss=$\textbf{9.0\%}&
$loss=$\textbf{43.9\%}&
$loss=$\textbf{3.7\%}&\textbf{16.1\%}\\

  \cline{1-6}  
$\mathcal{R}$ if Vela Jr. is 700 yrs old&
6.1\%(18.7\%):&
5.1\%(11.4\%):&
4.3\%(9.2\%):&
3.0\%(4.5\%):&\\

but prior assumes 4300 yrs of age&
$loss=$\textbf{67.4\%}&
$loss=$\textbf{55.1\%}&
$loss=$\textbf{53.5\%}&
$loss=$\textbf{33.3\%}&\textbf{52.3\%}\\

    \tableline
  \end{tabular}

\end{centering}
    \begin{tablenotes}
      \small
      \item\textbf{Notes:} The first number in each table cell is the $\mathcal{R}$ assuming a mismatch between the prior used in the optimization and the true age of the object. The number in parenthesis is the $\mathcal{R}$ if the prior is matched to the age of the object. The first number is always smaller than the $\mathcal{R}$ in parenthesis and the ratio in detection probability is in bold font.
    \end{tablenotes}
\end{threeparttable}

\end{table*}
From these it is clear that using the youngest-age prior for Vela Jr.  leads to the smallest loss in detection probability if this assumption is wrong, hence we use this prior in our optimizations. 

\subsection{The total computing budget}
\label{sec:totalBudget}

\begin{figure}
\centering
\includegraphics[width=0.45\textwidth]{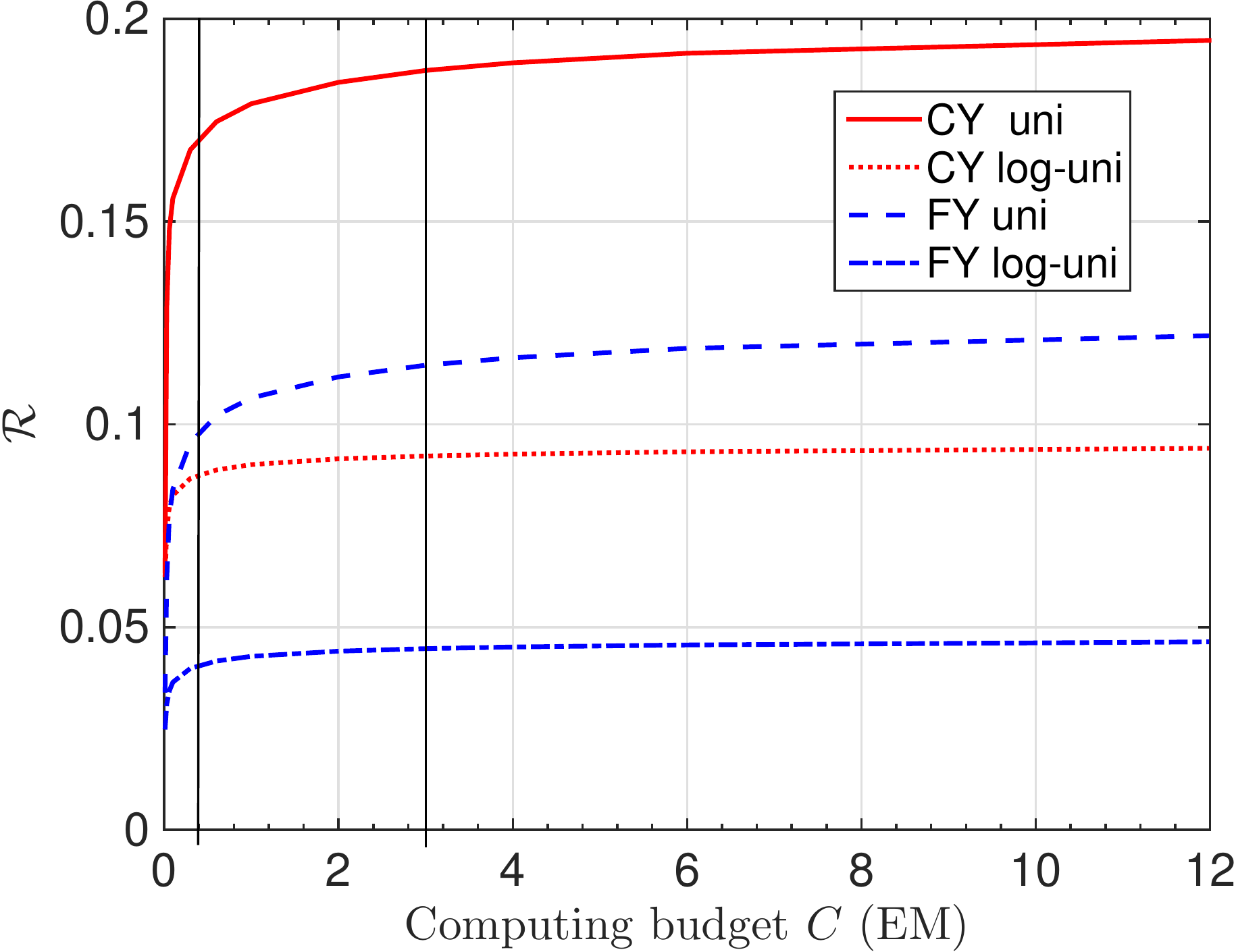}
\caption{$\mathcal{R}$ versus computing budget $C$}
 \label{c_vs_p}
\end{figure}

In the previous discussion we found that, if the likelihood of Vela Jr.'s frequency and spindown  is uniform between 20 and 1500 Hz, and if Vela Jr. is 700 yrs old and at a distance of 200 pc,  by optimally choosing set-ups and target parameter space to search with 3 EMs, the total detection probability $\mathcal{R}$ is 18.7\%. It is then natural to ask whether by investing more computing
resources we could achieve an even higher detection probability. Assuming a set of computing budgets from 0.1 EM to 12 EM, the optimization procedure yields the $\mathcal{R}$s shown in Fig.~\ref{c_vs_p} 
as a function of $C$. Although $\mathcal{R}$ always grows as $C$ increases, the growth rate decreases as $\mathcal{R}$ increases. We identify three stages. In the first stage, when $C$ is from 0.1 to 0.4 EM,  $\mathcal{R}$ increases very fast. In the second stage when $C$ is from 0.4 to $\sim$3 EM,  $\mathcal{R}$ still increases but not as fast as before. In the last stage when $C$ is larger than $\sim$3 EM, $\mathcal{R}$ increases even more slowly. In this regime a gain in the $\mathcal{R}$ due to 9 additional EMs, is less than 1\%. Based on this we decide to invest around 3 EM in this search, covering the first two stages of $C(\mathcal{R})$.

\subsection{Simplifications}
\label{subsec:simply}
%As discussed in Section \ref{subsec:PS} we optimized among 71 set-ups corresponding to five different values of $T_{\mathrm{coh}}$: 10 days, 15 days, 20 days, 30 days and 60 days. The optimal choice of set-ups for Vela Jr under the CY and uniform priors is shown in Fig.~\ref{fig:cyco}(a), (c) and Fig.~\ref{spdetail}(c) where we can see that 2 different set-ups for $T_{\mathrm{coh}}=10$ days, 7 set-ups for $T_{\mathrm{coh}}=15$ days, and  5 set-ups for $T_{\mathrm{coh}}=20D$ days have been picked just for Vela Jr.

As shown in the previous Sections, the optimal search set-up may well comprise different coherent time baselines $T_{\mathrm{coh}}$ in different $f-\dot{f}$ ranges for each target, and different grid spacings for the same $T_{\mathrm{coh}}$. For instance, under the CY and uniform priors assumption for Vela Jr., the set-prescription comprises 2 different set-ups for $T_{\mathrm{coh}}=10$ days, 7 set-ups for $T_{\mathrm{coh}}=15$ days, and  5 set-ups for $T_{\mathrm{coh}}=20$ days for Vela Jr.; 11 set-ups for $T_{\mathrm{coh}}=10$ days for Cas A and 27 set-ups distributed over four different $T_{\mathrm{coh}}$ values (10, 15, 20, 30 days) for G347.3. 

The analysis of the results from a search comprising this diversity in set-ups and coherent time baselines is quite daunting. So we examine the following question: how much detection probability would be lost if i) we considered only a single set-up per target ii) we extended the search frequency range for all targets to be fixed between 20-1500 Hz?  

\begin{figure*}%
    \centering
    \subfloat[Close and young, uniform distribution on $f$ and $\dot{f}$]{{  \includegraphics[width=.45\linewidth]{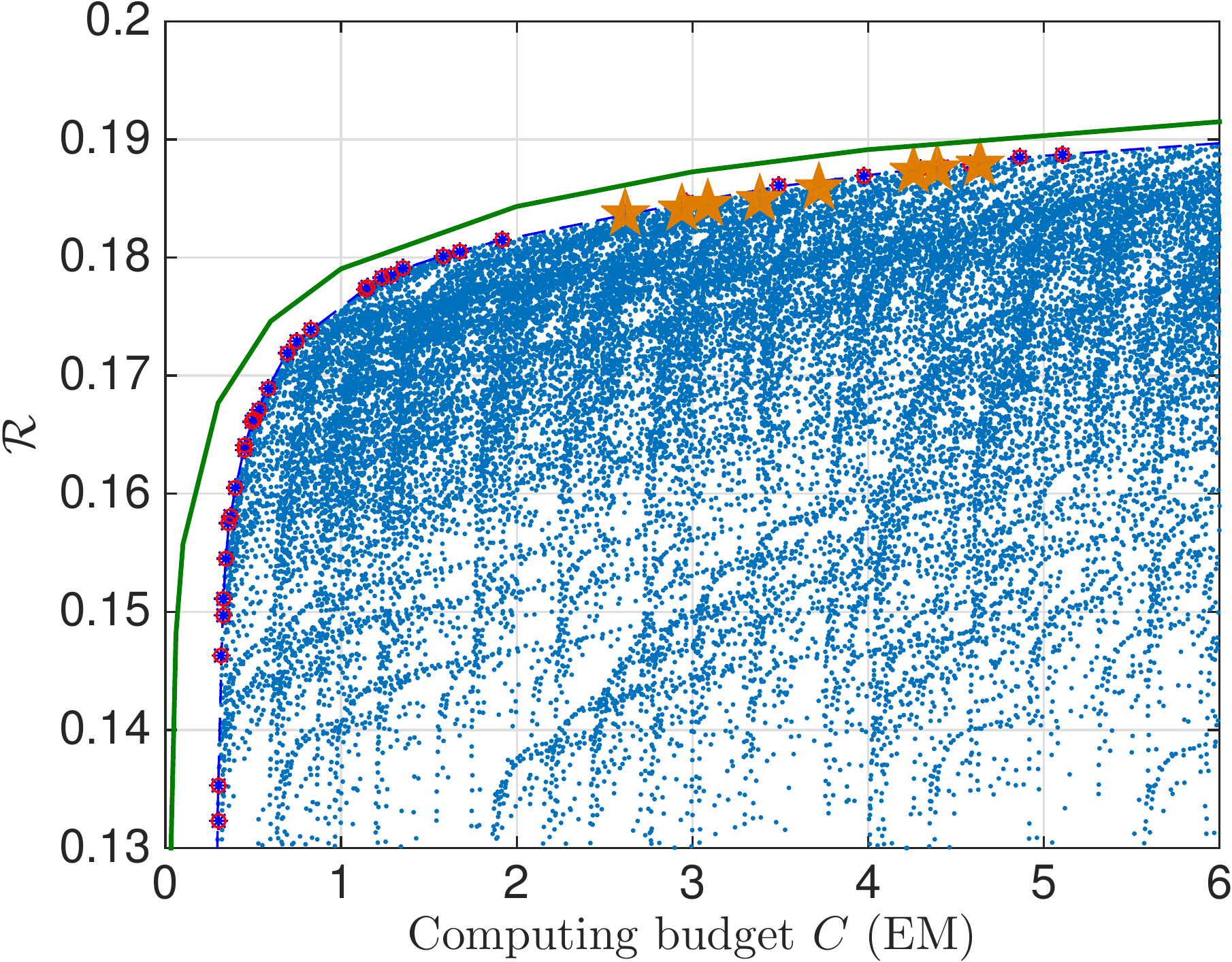}}}%
    \qquad
    \subfloat[Close and young, log-uniform distribution on $f$ and $\dot{f}$]{{  \includegraphics[width=.45\linewidth]{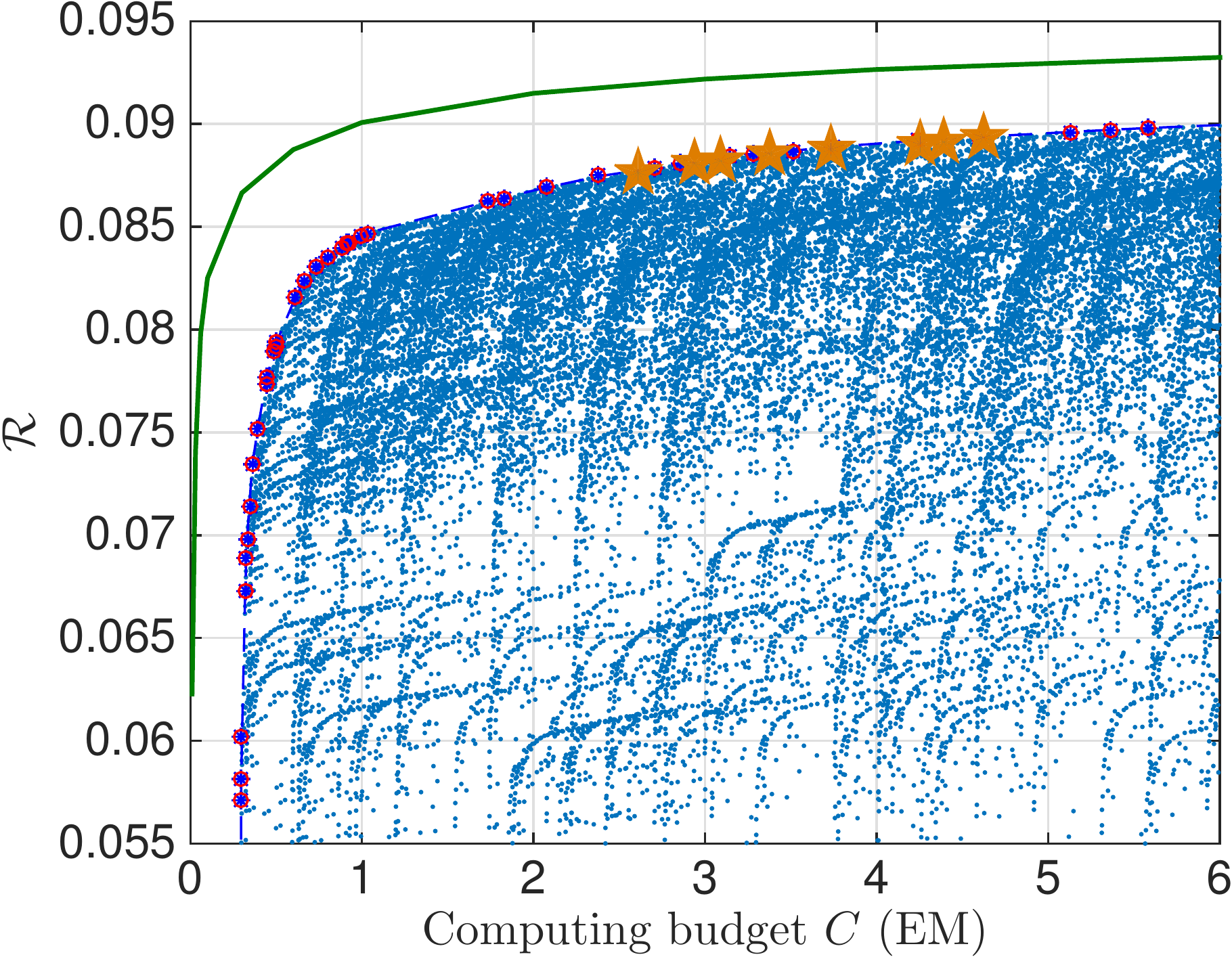}}}%
    \qquad
    \subfloat[Far and young, uniform distribution on $f$ and $\dot{f}$]{{  \includegraphics[width=.45\linewidth]{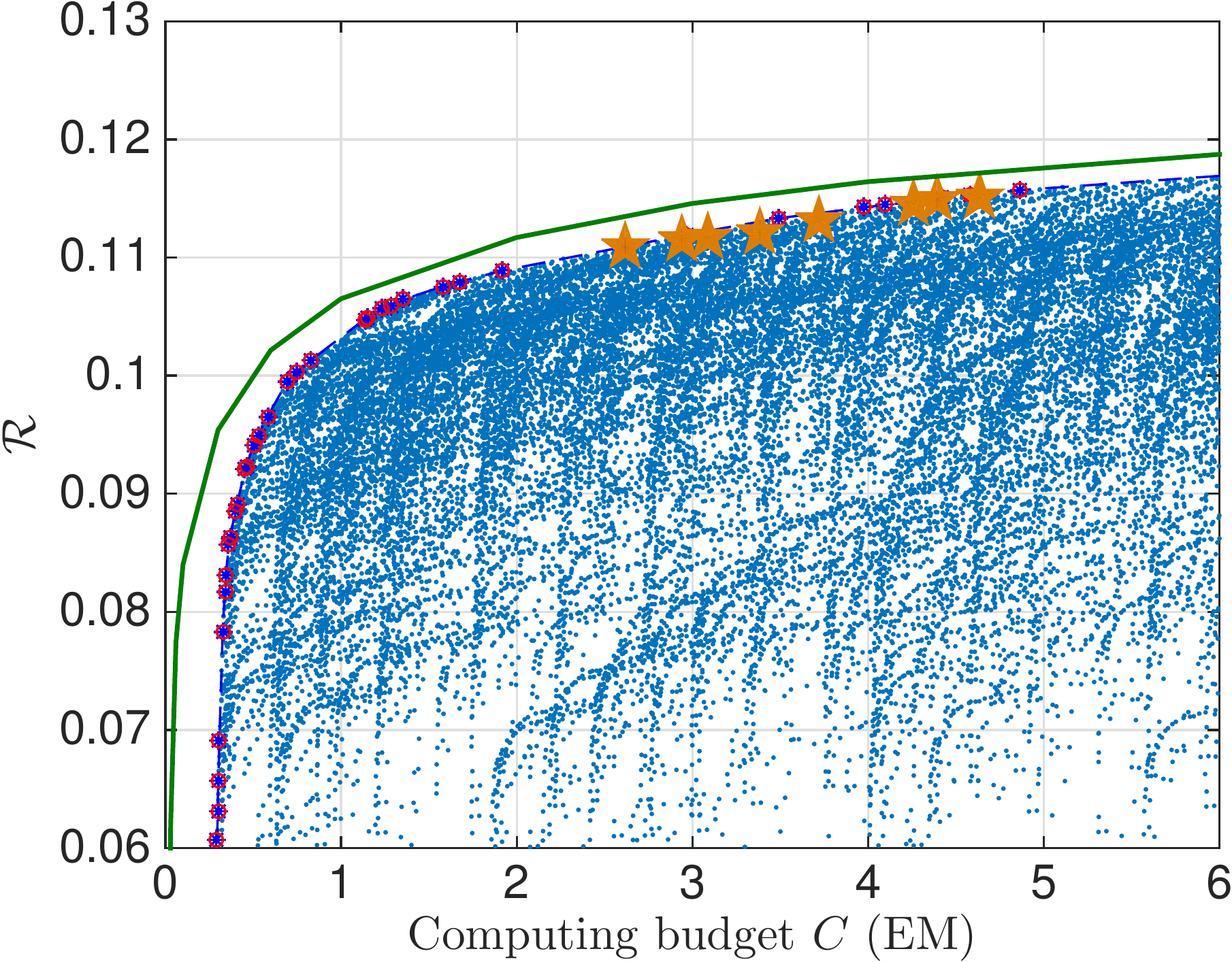}}}%
    \qquad
    \subfloat[Far and young, log-uniform distribution on $f$ and $\dot{f}$]{{  \includegraphics[width=.45\linewidth]{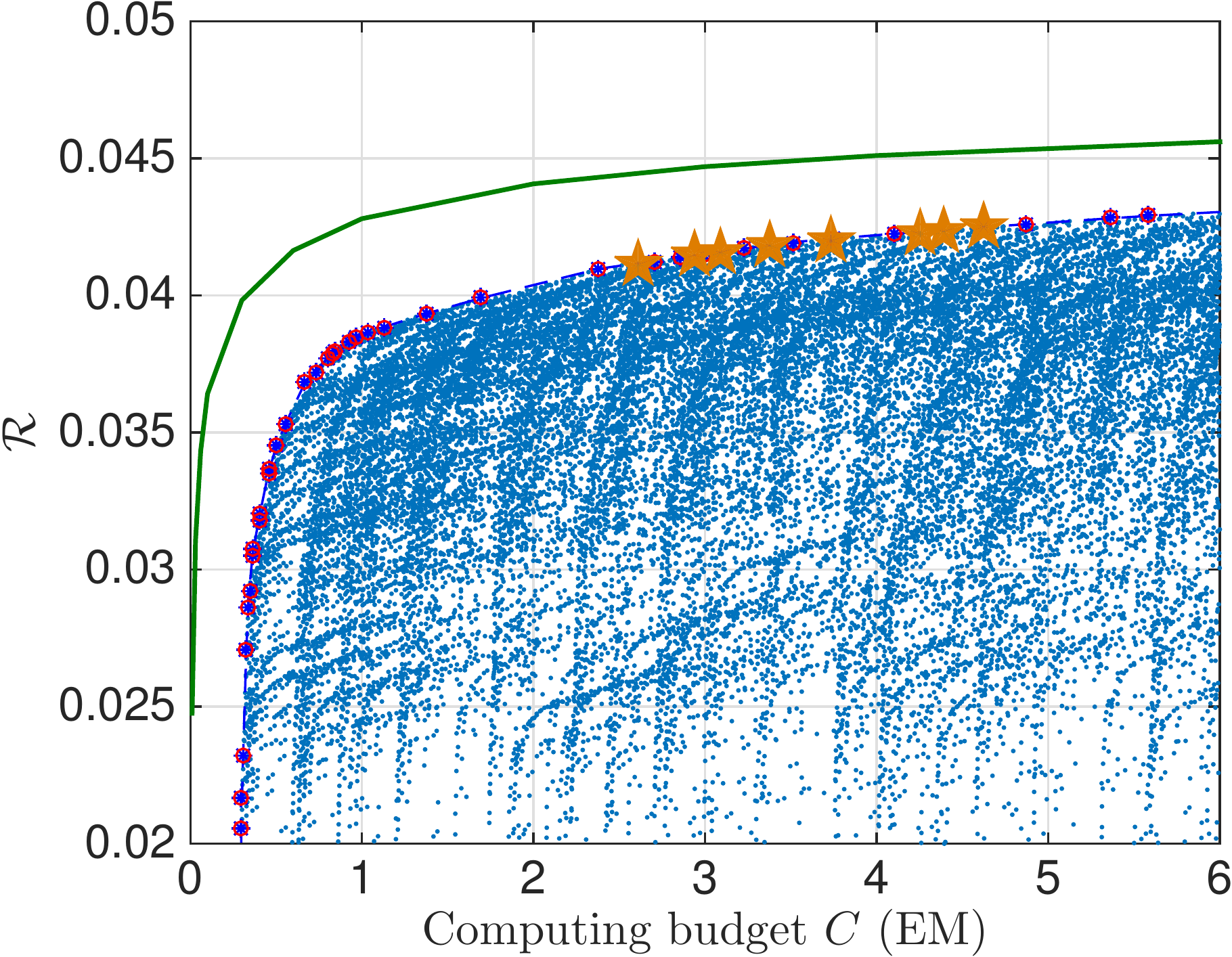}}}%
    \caption{$\mathcal{R}$ versus computing budget $C$ from 357911 combinations of setups with different priors.  Each dot on this plot is a combination of setups, one per target. The green curve shows the optimal $\mathcal{R}$ as a function of $C$. The golden stars highlight the best setup combinations around 3 EM.}%
    \label{Simplification_cy_uni}%
\end{figure*}

Limiting the set-up for each target to the 71 set-ups considered in Section \ref{subsec:PS} we have $71^3=357911$ combinations of setups. The computing costs of these ranges from a few 0.1EM to a few hundreds EM. We want a computing budget of a few EM. In Fig.~\ref{Simplification_cy_uni} we zoom in the cost range from 0 to 6EM and determine the total detection probability for all the set-up combinations. 

We find that the loss due to having restricted the choice to a single set-up per target is smaller than $\sim$0.3\% across all priors. Since this loss is much smaller than the total detection probability, we are persuaded to adopt this simplification and save ourselves a great deal of effort in the results post-processing phase. 

We pick a computing budget just below 5 EM and hence the set-ups corresponding to the right-most golden star of Fig.~\ref{Simplification_cy_uni}. Actually, all these golden stars correspond to the same set-ups independently of the priors for both the distance of Vela Jr. and $f$ and $\dot{f}$. The details of the chosen set-up are listed in Table~\ref{tab:finalsetup}. 
%{\mapcomment{why is this ? actually the set ups at the edges are not exactly the same. it's just that very close to the edges you can find the same set-ups, and these are the golden stars. The reason for this is that we have also constrained the search freq band, and so now the effect of unif versus log unif freq does not have a way to express itself in a different distribution of the comouting power. It does, though, show in the distance to the optimal detection probability.}} 
\begin{table}
\begin{threeparttable}
\caption{
  \label{tab:finalsetup}
Profile of the chosen set-ups }
  \begin{centering}
  \begin{tabular}[c]{c|cccccc|c}
    \tableline
   
Targets&$m_f$&$m_{\dot{f}}$&$m_{\ddot{f}}$&$\gamma^{(1)}$&$\gamma^{(2)}$&$T_\mathrm{coh}$&$\langle \mu\rangle$
\\
 \tableline
    
Cas A&0.3&0.5&0.003&4&20&10D&41.2\%\\
Vela Jr.&0.15&0.3&0.003&8&20&15D&15.8\%\\
G347.3&0.15&0.2&0.003&8&10&20D&12.1\%\\
 \tableline
  \end{tabular}

\end{centering}
    %\begin{tablenotes}
      %\small
      %\item\textbf{Notes:} 
    %\end{tablenotes}
\end{threeparttable}

\end{table}

We use arrows to indicate the three chosen set-ups in Fig.~\ref{spDetail}. Under the CY and uniform $f,\dot{f}$ assumption for Vela Jr., the chosen set-up for G347.3 is also the dominant one from the optimization scheme; the chosen set-up for Cas A is the second dominant one from the optimization scheme. Although the  chosen set-up for Vela Jr. is the fourth  dominant one from the optimization procedure (see Fig.~\ref{spDetail}(c)) the chosen $T_\mathrm{coh}=$  15 days is the same that of the dominant set-up from the optimization scheme. These differences can be explained considering that we take different prior combinations into account overall and the computing budget set for Fig.~\ref{spDetail} is 3 EM which is below the final-set budget 5 EM.

%??????????????????????
%{\mapcomment{{Are we taking uniform or log uniform priors for f and fdot and why ?}}

\section{Conclusions}
\label{sec:conclusions}

Following the search optimization procedure proposed in \cite{OptimalMethod}, we design a search using a few months of Einstein@Home optimized for a data set like the LIGO O1 data. We concentrate on 3 targets: Vela Jr., Cas A and G347.3. We extend the method proposed in \cite{OptimalMethod} by adding more dimensions to the optimization : i.e. we consider  different search set-ups for the same coherent time baseline $T_{\mathrm{coh}}$, varying the template banks in frequency and spindown, and we fold-in the {\it measured} mismatch distributions from the different banks. 

We also investigate how a mistake in choice of the astrophysical prior on the age of the target that contributes the most to the detection probability (Vela Jr.) would impact the detection probability, and then pick the prior that minimizes the loss.  

We study the dependency of the attainable detection probability on the computing budget, and, within practical constraints from running the search, we make sure that we have nearly saturated the detection probability growth. We pick a computing budget of $\sim$ 5 EM. 

After having obtained the optimal combination of set-ups for the different targets in the different regions of parameter space, we significantly simplify it in order to make the post-processing of the results less cumbersome. Even limiting the search set-up to a single one per target, we are able to achieve this without significant degradation in the detection probability. One may wonder if this doesn't prove that the optimization scheme is actually not very important. In a sense it does, at least for this data set. However, without knowing what the optimal is, we would not have been able to judge the goodness of any empirically motivated set-up. 

This is the final set-up chosen:
\begin{itemize}
\item For the youngest source Cas A, a set-up with 10 days coherent time baseline (12 segments) will be used. The computing cost employed on searching for a signal from Cas A is 1.7 EM. The detection probability is 1.2\%, if we assume uniform priors in frequency and spin-down; the detection probability is 0.2\%,  if we assume log-uniform priors.
\item For the closest  source Vela Jr., a set-up with 15 days coherent time baseline  (8 segments) will be used. The computing cost employed on searching for a signal from  Vela Jr. is 2.2 EM.  If we assume uniform priors in frequency and spin-down, a distance of 200 pc and an age of 700 yrs, the detection probability is 14.5\%. 
The detection probability drops to 3.8\% if we assume a distance of 750 pc and an age of 4300 yrs.
If we assume log-uniform priors in frequency and spin-down, a distance of 200 pc and an age of 700 yrs, the detection probability is 7.4\%. 
The detection probability drops to 3.0\% if we assume a distance of 750 pc and an age of 4300 yrs.

\item  For the second closest source G347.3, a set-up with 20 days coherent time baseline (6 segments) will be used. The computing cost employed on searching for a signal from Vela Jr. is 0.7 EM. The detection probability is 3.1\%, if we assume uniform priors in frequency and spin-down;  the detection probability is 1.4\%,  if we assume log-uniform priors.
\end{itemize}

\begin{figure}
\centering
\includegraphics[width=0.45\textwidth]{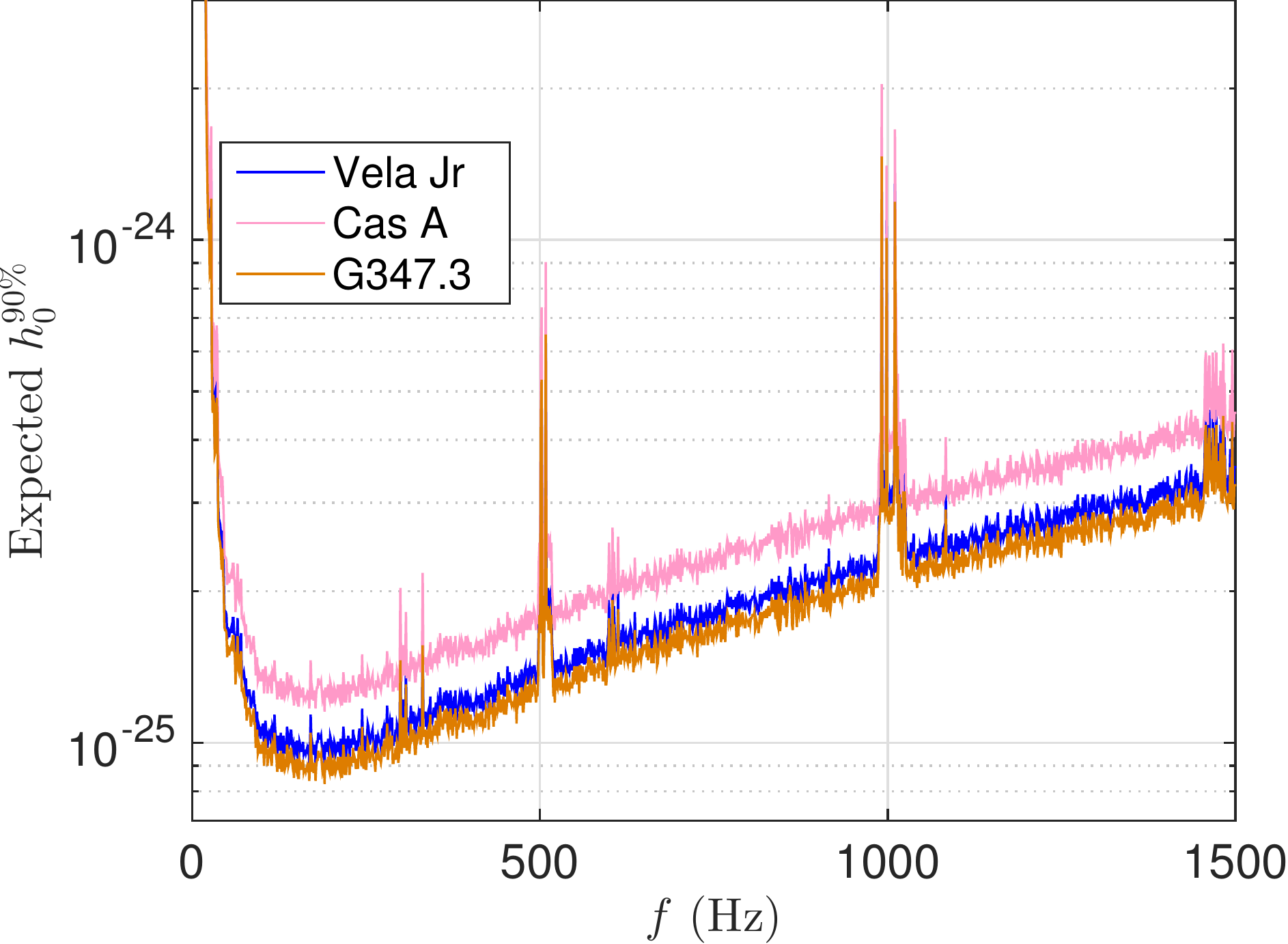}
\caption{Expected 90\% confidence upper limit on the GW amplitude $h_{0}^{90\%}$ from the directed searches proposed here, based on the noise level of the LIGO O1 data and on the estimated sensitivity of the proposed searches. We stress that this is {\it not} an observational result but a prediction of it.}
%{\mapcomment{remove empty space above 3e-24 and below 7e-26}}}
 \label{h0_90}
\end{figure}

The search that we propose here is a deep and broad frequency search
for all the three targets Vela Jr., Cas A and G347.3. It was set-up and ran for a few months on the volunteer computing project Einstein@Home during the first half of 2017. The post-processing of the results will be reported in a separate paper. In case no signal is detected the expected 90\% confidence upper limit on the GW
strain amplitude $h_{0}^{90\%}$ is shown in Fig.~\ref{h0_90}. At the
detector's most sensitive frequencies, $\approx $150 Hz,
$h_{0}^{90\%}$ for Cas A is $1.4\times10^{-25}$, for Vela Jr.
$1.0\times10^{-25}$ and for G347.3 $9\times10^{-26}$. We note that
$h_{0}^{90\%}=1.4\times10^{-25}$ for Cas A is 2 times smaller than the
S6 upper limit \cite{S6CasA} and deep searches on Vela Jr. and G347.3
have never been done yet. This can be quantified by using the notion
of sensitivity depth defined in \cite{Behnke:2014tma}.  The
sensitivity depth $\mathcal{D}^{90\%}$ of this search is
61.5~$\mathrm{Hz^{-1/2}}$ for Cas A, 79.1~$\mathrm{Hz^{-1/2}}$ for
Vela Jr., and 85.8~$\mathrm{Hz^{-1/2}}$ for G347.3. 
%\repradd{[comment: at what false-alarm level, or equivalently, at what threshold?]}

\section{Acknowledgements}

J.M. acknowledges support by the IMPRS on Gravitational Wave Astronomy
at the Max Planck Institute for Gravitational Physics in
Hannover. M.A.P. gratefully acknowledges support from NSF PHY grant
1104902. We thank David Keitel, Ra Inta and Ben Owen for valuable
comments.  This paper was assigned LIGO document number P1700166.

\bibliography{refs}

\end{document}